\documentclass[twocolumn]{aastex701}

\usepackage{tabularx}
\usepackage{amsmath}
\usepackage{graphicx}
\usepackage{booktabs}
\usepackage{multirow}
\usepackage{accents}
\usepackage{tablefootnote}
\let\tablenum\relax 
\usepackage{siunitx}
\usepackage{needspace}  \Needspace{5\baselineskip}
\usepackage{placeins}
\usepackage{longtable}
\usepackage{amssymb,mathtools,bm}
\usepackage{booktabs}
\usepackage{enumitem}
\usepackage{xcolor}

\newcommand{\V}{\mathbf{V}}

\newcommand{\vesc}{v_{\rm esc}}

\newcommand{\p}{\partial}
\newcommand{\dvpar}{\Delta v_{||}}
\newcommand{\dvper}{\Delta v_{\perp}}
\newcommand{\dvv}{\Delta v_V}
\newcommand{\dvb}{\Delta v_b}
\newcommand{\dv}{\Delta v}
\newcommand{\dE}{\Delta E}
\newcommand{\dJ}{\Delta J}
\newcommand{\Mbh}{M_\bullet}

\usepackage{accents}

\shortauthors{Distler and Stone}
\shorttitle{Strong Scatterings in Dense Star Clusters}
\begin{document}
\title{Ejected and Rejected: Strong Scatterings in Dense Star Clusters}

\author[orcid=0009-0006-4294-6760]{Adam Distler}
\altaffiliation{Fannie and John Hertz Foundation Fellow}
\altaffiliation{NSF Graduate Research Fellow}
\affiliation{Center for Astrophysics | Harvard and Smithsonian, 60 Garden Street, Cambridge, MA 02138, USA}
\affiliation{Department of Astronomy, University of Wisconsin-Madison, 475 N.~Charter St., Madison, WI 53706, USA}
\email[show]{adam.distler@cfa.harvard.edu}
\author[orcid=0000-0002-4337-9458]{Nicholas C. Stone}
\affiliation{Department of Astronomy, University of Wisconsin-Madison, 475 N.~Charter St., Madison, WI 53706, USA}
\affiliation{Racah Institute of Physics, The Hebrew University, 91904, Jerusalem, Israel}
\email{nstone2@wisc.edu}
\begin{abstract}
Understanding the evolution of collisional stellar systems is key to predicting both the bulk evolution of dense star clusters and also the rates of tidal disruption events, extreme mass ratio inspirals (EMRIs), stellar ejections, and other loss cone phenomena. Although a majority of the evolution is dictated by weak, small-angle scatterings, rare close encounters resulting in strong scatterings can sometimes produce qualitatively different outcomes. To systematically determine the astrophysical importance of strong scatterings, we derive the exact differential rate distribution of velocity kicks in isotropic and spherically symmetric distribution functions representing globular and nuclear star clusters, while also deriving the distribution boundaries and approximate rate scalings. We then extend this formalism to derive approximate rates of energy and angular momentum kicks for objects in the nuclear star cluster, and then apply this framework to study (i) ejections and (ii) ``cancellation'' of extreme mass ratio inspirals. We find that ejection rates are complicated in some regimes due to the saturation of the strong kick distribution and the breakdown of the weak limit, likely enhancing the effect of loss cone shielding. We also determine that a sizable fraction of proto-EMRIs will likely be canceled by strong scatterings, highlighting the need for future work to determine to what degree strong scatterings impact the true EMRI rate.

\end{abstract}

\keywords{Astrodynamics (76), Celestial mechanics (211), Gravitation (661), Close encounters (255)}

\section{Introduction}\label{sec:intro}
The evolution of many-body collisional star systems is a foundational problem in modern astrophysics: most stars are born in the collisional environment of a gas-embedded cluster \citep{LadaLada2003}, a substantial minority pass through an open cluster environment, and even denser systems like globular (GCs) and nuclear star clusters (NSCs) are responsible for the production of unique high-energy sources and transients \citep{Clark1975, Hills1975, Ryan1995, PortegiesZwart2000, Generozov2018, Miniutti2019, Kirsten2022}.  The dynamical evolution of dense star systems can be studied in a first-principles way using direct $N$-body simulations.  However, because the cost of these simulations often becomes prohibitive for $N\gg 1$, techniques from statistical mechanics can be more directly useful. In this picture, the background population of stars is idealized as a distribution function (DF), and its fluctuations and discreteness can be treated statistically.

One of the most fundamental tools for modeling the evolution of many-body systems is the calculation of diffusion coefficients \citep[DCs;][]{Chandrasekhar1943, BinneyTremaine}, which most often represent the statistical outcomes of many weak, pair-wise two-body scatterings\footnote{Other kinds of DCs can be computed to represent the statistical outcomes of many other pair-wise interactions, such as secular torques \citep{BarOrFouvry18}.}. These DCs are calculated from a rate-weighted average of the underlying distribution of velocity kicks, itself determined from the DF.  Once computed, DCs can be used in different evolutionary frameworks: for example, the Fokker-Planck equation uses DCs to evolve the entire test star DF, while the Langevin equation merely evolves the orbit of a single test star.  In these frameworks, the first and second-order DCs dictate deterministic drift and stochastic diffusion, respectively. 

The two-body drift coefficient was first used in astrophysical environments to identify the dynamical friction force \citep{Chandrasekhar1943}, which has since been understood to play a major role across many scales of astrophysical dynamics, from mass segregation in globulars \citep{Spitzer1971} and the Milky Way's Galactic Center \citep{Morris1993}, to the infall of satellite galaxies into their primary hosts \citep{OstrikerTremaine75}. Likewise, the second-order diffusion coefficients are the drivers of other important astrophysical phenomena, such as core collapse of globulars and the production of loss cone transients \citep{Merritt2013b}.  Taken as a set, the DCs are used to calculate the evolution of the DF through the Fokker-Planck equation, for example in the search for steady state, multi-mass solutions \citep{BW1976, BW1977, AlexanderHopman2009, Linial2022}. Individual particle orbits can also be simulated through the use of the Langevin equation \citep{Shapiro1978}, which is particularly useful for investigating the properties of extreme mass ratio inspirals (EMRIs) in nuclear star clusters \citep[NSCs;][]{2005ApJ...629..362H, Qunbar_stone, Mancieri2025}.


Despite their evident scientific value, these approaches cannot capture strong gravitational encounters that result in non-diffusive motion through phase space. Although usually rare, a single strong encounter can qualitatively alter the evolution of a particle by sending it to parameter space not immediately accessible to the first and second-order DCs. Previous works focusing on strong scatterings have investigated properties of the underlying kick distributions \citep{Henon1969, Goodman1983}. This kick distribution can yield further insights by conveniently providing higher-order DCs \citep{Bar-or2013}, estimating rates of cluster evaporation from ejections \citep{LinTremaine1980}, and identifying novel phenomena, such as loss cone shielding \citep{Teboul2024, Kaur2024, TeboulPerets2025}.

Notably, previous works on strong scatterings have focused either on rates of specific phenomena such as ejections (e.g. \citealt{LinTremaine1980, Teboul2024}) or on a spectrum of outcomes as is quantified by transition probability operators between different states (energy, for example). However, the calculation of this operator is typically complex, which both limits the direct intuition it provides and hampers its extension into higher dimensions (for example, we are unaware of the calculation of a strong scattering transition probability operator in energy {\it and} angular momentum space). In this paper, we instead focus our effort on deriving the differential rate distribution of strong velocity kicks.  
This differential rate distribution contains the same information content (and quantifies the same dynamical physics) as a transition probability operator \citep{Goodman1983}, but as we shall see, it can be written in closed form in two-dimensional velocity space.  The differential rate distribution readily yields higher moments beyond the first and second-order DCs, and upon specifying the geometry of the stellar system, it can also be converted to obtain the rate distribution of strong kicks to energy and angular momentum (AM), yielding further insights into ejections and gravitational wave phenomena.

The work is organized as follows: Section~\ref{sec:the_DRD} outlines the analytic derivation of the two-dimensional differential rate distribution, along with concrete calculations for globular clusters (Section \ref{sec:globular}) and NSCs (Section \ref{sec:NSC}). Section~\ref{sec:third_mom} outlines a discussion of the third moment's calculation and approximations for the NSC case; Section~\ref{sec:action_applications} outlines and approximates the recasting of the rate distribution into action space, and introduces applications to EMRI cancellation and generic ejections.  Section~\ref{sec:conclusion} summarizes this work and discusses further applications and extensions.

\section{The Differential Rate Distribution}\label{sec:the_DRD}
We consider the interaction between a test object of mass $M$ that undergoes a hyperbolic encounter with a field star of mass $M_{\rm s}$. The encounter possesses a relative velocity $V = |\Vec{V}|=|\Vec{v}-\Vec{v_f}|$ (with $\vec{v_f}$ being the field star velocity and $\vec{v}$ being the test star velocity, both in the lab frame) and impact parameter $b=|\Vec{b}|$. We treat this encounter as inherently local, unaffected by the presence of a broader GC or NSC potential during the scattering. The changes to the velocity of the test star after this encounter can be written \citep{Merritt2013}
\begin{equation}
    \Delta \Vec{v} =  \Delta v_V \hat{V} + \Delta v_b \hat{b}\,, \label{eq:dv_enc}
\end{equation}
with a hat denoting a unit vector $\hat{V} = \Vec{V}/|V|$ and
\begin{equation}
    \begin{gathered}
        \Delta v_V = \frac{-2\mu V}{1+ (b/b_0)^2}, \\
        \Delta v_b = \frac{-2\mu V (b/b_0)}{1+ (b/b_0)^2}\,.
    \end{gathered} \label{eq:dvv_dvb_defs}
\end{equation}
Here, $\mu = M_{\rm s}/(M+M_{\rm s})$, $b_0$ is the impact parameter $b_0 = G (M_{\rm s}+M)/V^2$ that results in a 90-degree deflection, and $G$ is Newton's constant. The magnitude of $\Delta \Vec{v}$ is invariant under rotational and translational coordinate changes (see Appendix~\ref{sec:scat_geo} for more explicit vector geometry), and we can write that the magnitude of the change in velocity is 
\begin{equation}
    \Delta v^2 = \frac{4\mu^2 V^2}{1+(b/b_0)^2}\,.
\end{equation}

In the limit where $b/b_0$ is large, we can write that $b\approx 2G M_{\rm s}/ (V \Delta v)$. To construct the differential rate distribution (DRD) of velocity kicks, we estimate the encounter rate $\Gamma$ using the $n-\Sigma-V$ approximation, 
which is generally valid in collisional star systems \citep{Modelevsky2025}.  We will begin by calculating a very approximate form of the DRD to build intuition, and afterwards we will estimate the DRD more rigorously.  First, we will assume a constant number density of field stars $n_*$ with a uniform relative velocity. The differential rate can then be written
\begin{equation}
    \frac{{\rm d}\Gamma}{{ \rm d}b} =  2\pi b V n_*\,, \label{eq:DRDapprox}
\end{equation}
and in the limit of large $b/b_0$,
\begin{equation}
    \frac{{\rm d}\Gamma}{{\rm d}\Delta v} = \frac{8\pi G^2 M_{\rm s}^2 n_*}{V \Delta v^3} \label{eq:twiddle_dist}\,.
\end{equation}
While the overall scaling of ${\rm d}\Gamma/{\rm d}\Delta v \propto \Delta v^{-3}$ is evident, we are more specifically interested in the differential rate distribution as a function of changes to the test object's original direction of motion ($\dvpar$), and the motion perpendicular to this ($\dvper$). Defining the lab frame such that the object's original velocity is parallel to the $x$-axis, we can also identify a generic boundary to the kick distribution. From Eq.~\ref{eq:dv_enc} and Eq.~\ref{eq:dvv_dvb_defs}, a dot product of the velocity kick vector with the relative velocity vector yields
\begin{equation}
    \Delta \Vec{v} \cdot \Vec{V}= \frac{-2\mu V^2}{ 1+(b/b_0)^2} = -\frac{\dv^2}{2\mu}\,. \label{eq:scat_iden1}
\end{equation}
Upon expanding $\Vec{V} = \Vec{v} - \Vec{v_f}$,
\begin{equation}
    \Delta \Vec{v} \cdot \vec{v_f} = v \dvpar +\frac{\dv^2}{2\mu}\,.
\end{equation}
In systems such as globulars or NSCs, there is a maximum velocity a particle can possess before it escapes the system, defined as $\vesc$. By the Cauchy-Schwarz inequality and bounding $|\Vec{v_f}|\le v_\mathrm{esc}$ for the generic escape velocity of a system, we have that
\begin{equation}
     |\kappa| \equiv \left|\frac{v \dvpar}{\dv} +\frac{\dv}{2\mu} \right|\le v_\mathrm{esc}\,.
\end{equation}
At the boundary, we have equality of $|\kappa|=\vesc$; squaring both sides yields the following equation for the maximum kicks  $\dvpar$ and $\dvper$ that can be received: 
\begin{equation}
     \frac{(\dvpar^2 + \dvper^2)}{4\mu^2} +v\frac{\dvpar}{\mu} + v^2 \frac{\dvpar^2}{\dvpar^2 + \dvper^2}  = v_\mathrm{esc}^2  \label{eq:kick_boundary}
\end{equation}

\begin{figure}
    \centering
    \includegraphics[width=0.97\linewidth]{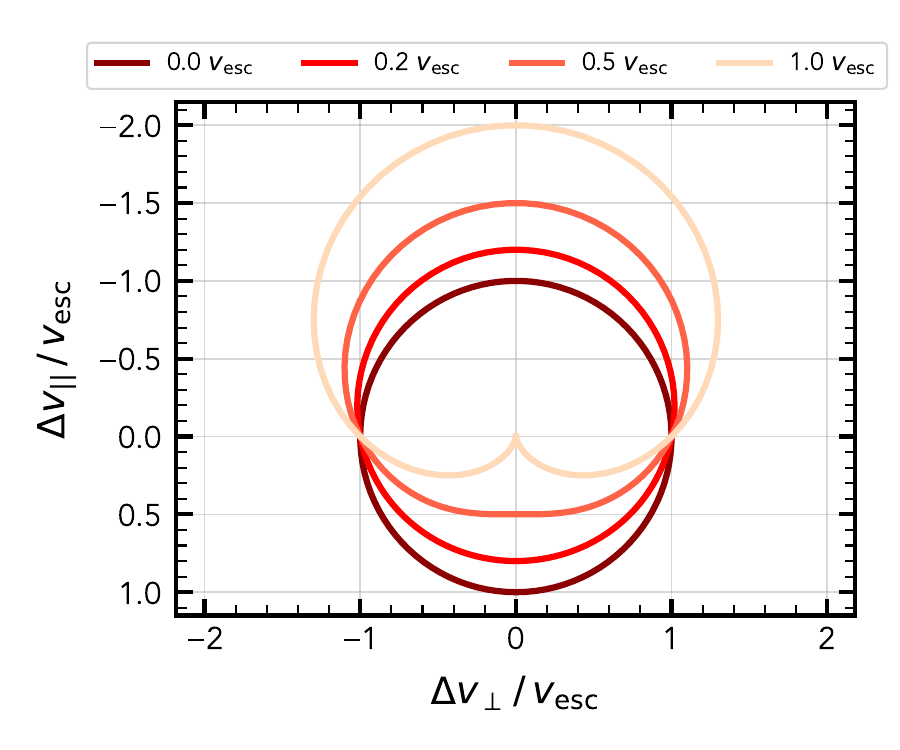}
    \caption{Kick boundary for strong scatterings for $M=M_{\rm s}$ (Eq. \ref{eq:kick_boundary}). The darker colors correspond to slower velocities, ranging from $v=0$ to $v=v_\mathrm{esc}$. We have mirrored $\dvper$ for visualization purposes, although by definition $\dvper>0$. The kick boundary is symmetric for $v=0$ (reflecting the isotropy of field star velocities) but becomes increasingly biased towards $\dvpar$ as $v\to v_{\rm esc}$: the test star experiences a headwind of field stars in this limit, and preferentially slows down.}
    \label{fig:kick_bounds}
\end{figure}

A figure displaying the changing kick boundary is shown in Figure~\ref{fig:kick_bounds} for a $M=M_{\rm s}=1\,M_\odot$ object orbiting a $10^6\,M_\odot$ supermassive black hole (SMBH), assuming a Keplerian potential. We see that for $v=0$, the kick boundary is symmetric in $\{\dvpar, \dvper \}$; however, a strong asymmetry emerges as the test particle approaches the escape velocity. In this limit, a strong headwind of particles allows for strong scatterings that can remove a sizable fraction of the test particle's velocity, or even reverse their instantaneous velocity, but that limits the opportunity for test particles to get ``boosted'' by strong scatterings along their current direction of motion. 

To determine the maximum allowed kick accounting for any scattering geometry, we introduce the parameterization $\dvpar = \dv \cos(\chi)$ and $\dvper = \dv \sin(\chi)$ with $0\le \chi \le \pi$. Substituting into Eq.~\ref{eq:kick_boundary} yields $\dv = 2\mu [v_\mathrm{esc} - v\cos(\chi)]$. To then find the corresponding maxima and minima for $\dvpar$ and $\dvper$, we can find extrema of $\dvpar(\chi)$ and $\dvper(\chi)$. We find that if $v < v_\mathrm{esc}/2$, then
\begin{equation}
     \Delta v_{||,\mathrm{max}} = 2\mu (v_\mathrm{esc}-v)\,,\hspace{0.3cm} \Delta v_{||,\mathrm{min}} = -2\mu (v_\mathrm{esc}+v)\,.
\end{equation}
Conversely, if $v>v_\mathrm{esc}/2$, then
\begin{equation}
 \Delta v_{||,\mathrm{max}} = \frac{ \mu v_\mathrm{esc}^2}{2v}\,, \hspace{0.3cm} \Delta v_{||,\mathrm{min}} = -2\mu(v_\mathrm{esc}+v)\,.  
\end{equation}

The corresponding perpendicular maxima are
\begin{equation}
\begin{gathered}
        \Delta v_{\perp,\mathrm{max}} =  2 \mu (v_\mathrm{esc}-v\lambda) \sqrt{1-\lambda^2}, \\
        \lambda = \frac{v_\mathrm{esc}- \sqrt{v_\mathrm{esc}^2 + 8 v^2}}{4 v}  .
\end{gathered}
\end{equation}

These bounds are generic, as they make no assumptions about the underlying field star velocity profile\footnote{A partial exception to this statement would emerge in extremely anisotropic systems which completely lack any field stars with velocities $v_{\rm f}$ above (or below) some fraction of the local $v_{\rm esc}$.}, number density, or (an)isotropy. Instead, the boundary is set by the maximum velocity a field star can possess, the velocity of our test object, and the masses of the two components of the scattering.

We now perform a more accurate calculation of the rate distribution by using a more generic version of the differential rate, i.e. by improving on Eq. \ref{eq:DRDapprox}. Assuming a spherical system and an isotropic DF, we have that
\begin{equation}
    {\rm d}\Gamma = b V f(v_f) \,{\rm d}b \,{\rm d}\psi \, {\rm d}^3 v_f\,, \label{eq:rate_measure}
\end{equation}
with $\psi$ a scattering angle (Appendix~\ref{sec:scat_geo}) and $f(v_f)$ the DF. For the stellar systems we consider, $f$ is a function of the distance to the origin of the system, but we omit writing $f(v_f, r)$ for brevity. The differential distribution of velocity kicks can then be written
\begin{equation}
\begin{gathered}
        \frac{{\rm d}^2\Gamma}{{\rm d}\dvpar {\rm d}\dvper} = \int {\rm d}b\, {\rm d}\psi \, {\rm d}^3 v_f \, b V f(v_f)\\
        \times \delta[ \dvpar -\accentset{\sim}{\dvpar }(b,V,\psi)] \delta[ \dvper -\accentset{\sim}{\dvper}(b,V,\psi)]\,,
\end{gathered}
\end{equation}
where the $\delta$-functions enforce the specific geometry yielding a given pair of observed kicks, $(\dvpar, \dvper)$. The overtilde labels the kicks $(\accentset{\sim}{\dvpar}, \accentset{\sim}{\dvper})$ given by a specific scattering geometry $(b,V,\psi)$. The equations for $\accentset{\sim}{\dvpar}(b,V,\psi)$ and $\accentset{\sim}{\dvper}(b,V,\psi)$ can be found in Eqs.~\ref{eq:dvpar_b_V_psi} and \ref{eq:dvper_b_V_psi}, respectively. Following substantial simplification (which we omit here but can be found in Appendix \ref{sec:drd_der}), the DRD integral can be written as
\begin{equation}
    \frac{{\rm d}^2\Gamma}{{\rm d}\dvpar {\rm d}\dvper }= \frac{ 8\pi G^2 M_{\rm s}^2 \dvper}{(\dvpar^2 + \dvper^2)^{5/2}} \int_{\Delta \hat{v}\cdot \Vec{v_f} = \kappa} f(v_f) dA \,.
\end{equation}
Here, $\Delta \hat{v}\cdot \Vec{v_f} = \kappa$ defines a plane in the space of field star velocities, with minimum distance $\kappa$ from the $\vec{v_f}$ origin. For a DF that can be written solely as a function of the specific energy $E$, we will find it helpful to define
\begin{equation}
    \xi^2 = v_f^2 - \kappa^2 \,.
\end{equation}
Then, the specific energy can be recast in terms of $\xi$ and $\kappa$
\begin{equation}
    E(\xi) = \Phi(r)-\frac{1}{2}( \kappa^2 + \xi^2)\,,
\end{equation}
where $\xi=0$ is the minimum distance from the plane to the origin and $\Phi(r)$ is the gravitational potential. We note that we are using the positive-bound convention of e.g. \citet{BinneyTremaine}, such that $\Phi>0$ and $E>0$ for bound orbits. Correspondingly, we can define
\begin{equation}
     E_* = \Phi(r)-\frac{1}{2}\left( \frac{\dv}{2\mu} + v \frac{\dvpar}{\dv}\right)^2\,.
\end{equation}
The disk ends where $E(\xi)=0$, or where $\xi=\sqrt{2E_*}$, since we can write that
\begin{equation}
    E(\xi) = E_* - \frac{\xi^2}{2}
\end{equation}
So then our rate distribution finally becomes
\begin{equation}
    \frac{{\rm d}^2\Gamma}{{\rm d}\dvpar {\rm d}\dvper }= \frac{ 16\pi^2 G^2 M_{\rm s}^2 \dvper}{(\dvpar^2 + \dvper^2)^{5/2}} \int_0^{\sqrt{2E_*}} \xi f(E_* - \frac{\xi^2}{2}) d\xi \,.
    \label{eq:master_drd_iso}
\end{equation}
Further simplification of the DRD is not possible without specifying a particular DF. We will now consider the two most relevant DFs for collisional star clusters.

\subsection{Globular Cluster - King Model}
\label{sec:globular}

Globular clusters are composed of $\sim 10^{4-6}$ stars and are abundant constituents of most galaxies.  Observations of Milky Way GCs show that they are, to a reasonable approximation, both spherical and isotropic, an observation which may be explicable given that their relaxation times are typically much less than the Hubble time \citep{king1968, spitzer_dyn_ev_gc}.

As globulars are dominated by the self-potential of their own stars, collisional relaxation drives the DF towards an isothermal (Maxwellian) distribution. However, due to tidal effects within the host galaxy, a GC will often truncate at a tidally limited radius, allowing stars to evaporate \citep{SpitzerHarm1958}. As a result, a lowered Maxwellian is a more realistic model for the DF of a globular cluster; a particular family of which is the King model \citep{King1966}, with the DF given by
\begin{equation}
    f = D (e^{E/\sigma^2}-1)\,.
\end{equation}
Here, $\sigma$ is the (constant) velocity dispersion and $D$ is a normalization constant related to the density\footnote{Naturally, in the regime where $E<0$, the DF is replaced by $f\equiv 0$.}. Combining Eq.~\ref{eq:master_drd_iso} with the DF of the King model, we find the globular DRD
\begin{equation}
    \frac{{\rm d}^2\Gamma}{{\rm d}\dvpar {\rm d}\dvper} =  \frac{16\pi^2 G^2 M_{\rm s}^2 D \dvper}{(\dvpar^2+\dvper^2)^{5/2}} [\sigma^2 e^{E_*/\sigma^2} - \sigma^2 - E_*]\,. \label{eq:DRDglobular}
\end{equation}
This equation can be used to analytically predict the rates of interesting astrophysical phenomena in globular clusters such as runaway star ejections \citep{Cabrera2023}, and possibly also collisions \citep{Glebbeek2008}; it may be particularly relevant for the buildup of a central very massive star via collisional runaway in a newborn GC. 

A substantial fraction of nuclear star clusters in dwarf galaxies seem to lack central SMBHs \citep{Neumayer2020}, and are thus also expected to relax into nearly isothermal spheres, with DRDs describable by Eq. \ref{eq:DRDglobular}.  For the remainder of this work, however, we focus our attention on NSCs dominated by the potential of a central SMBH.

\subsection{NSC - Power Law Density Profile}
\label{sec:NSC}
Consider a spherically symmetric and isotropic NSC with a central SMBH of mass $\Mbh$ and a density profile $\rho = \rho_\mathrm{inf} (r/r_\mathrm{inf})^{-\gamma}$ up to the radius of influence $r_\mathrm{inf}$. At $r=r_{\rm inf}$, where the enclosed mass equals that of $\Mbh$, the field star density is $\rho_\mathrm{inf}$.
An Eddington integral yields the DF $f(E) = C_\gamma E^{\gamma-3/2}$, with $C_\gamma$ being a normalization constant that is a function of the power-law index $\gamma$, which itself must lie below $3$ (at which point the enclosed mass diverges) and above $1/2$ \citep[at which point the velocity field can no longer be isotropic;][]{Tremaine1994, An2006}.

Many different density profiles for NSCs are plausible, as the complex processes of star formation and globular cluster infall (the two main formation pathways for NSCs) do not inherently need to produce power-laws in $\rho$.  Nonetheless, power-law density profiles {\it are} the natural expectation for NSCs that have had time to collisionally relax since their formation.  
This result was first seen in the seminal work of \citet{BW1976}, which found that $\gamma=7/4$ is a quasi-stationary\footnote{NSCs can never relax to a true stationary state because of the presence of the central SMBH, which will slowly consume the stellar population through the process of tidal disruption.  However, as long as the timescale to deplete stars through tidal disruption (or, as we shall later investigate, through ejection in strong scatterings) is long compared to the relaxation time, quasi-stationary states can be achieved.} state (``zero flux'') solution to the single-mass 1D Fokker-Planck equation in energy space.  Subsequent studies that incorporated a multi-mass model (which is crucial for systems with a large spread in mass, such as mixtures of stellar-mass BHs and lower-mass field stars) find that rarer heavy objects tend to form a steeper power-law cusp with slope $\gamma_{\rm H}$, and light objects form a shallower cusp with $\gamma_{\rm L}$, leading to mass segregation.  In the ``weak segregation'' regime, the heavy species dominates energy diffusion rates; here $\gamma_{\rm H}=7/4$ and $\gamma_{\rm L}=3/2$ \citep{BW1977}.  
In contrast, the ``strong segregation'' regime, in which the heavy species is rarer and thus experiences a more pronounced form of dynamical friction, remains a subject of theoretical controversy.  Early work found that the lighter stars maintain in this regime a shallow cusp of $\gamma_{\rm L} \approx 3/2\mathrm{-}7/4$, while the heavy objects attain extreme slopes, with $\gamma_{\rm H} \approx 7/4-11/4$ \citep{AlexanderHopman2009}.  The finite-flux strong segregation solutions of \citet{AlexanderHopman2009} cannot be self-consistent across all radii, however \citep{Linial2022}, and more recent work has explored zero-flux alternatives \citep{Linial2022}.  

For practical calculations, we take the one-dimensional galaxy-scale velocity dispersion $\sigma_\mathrm{BH}$ to be based on the empirical $M-\sigma$ relation \citep{Gultekin2009}
\begin{equation}
    \sigma_\mathrm{BH} = 70 \bigg(\frac{\Mbh}{1.53\times 10^6\,M_\odot}\bigg)^{1/4.24} \,\mathrm{km/s} \,.
\end{equation}
The influence radius can then be defined such that $r_\mathrm{inf} = G \Mbh (1+\gamma)^{-1}/ \sigma_\mathrm{BH}^2$. The local velocity dispersion at a radius $r<r_\mathrm{inf}$ is then $\sigma(r)^2 = G\Mbh (1+\gamma)^{-1} / r $. Applying Eq.~\ref{eq:master_drd_iso} to the NSC DF results in
\begin{equation}
 \frac{{\rm d}^2\Gamma}{{\rm d}\dvpar {\rm d}\dvper} = \frac{16 \pi^2 G^2 M_{\rm s}^2 C_\gamma}{\gamma-1/2} \frac{\dvper}{(\dvpar^2 + \dvper^2)^{5/2}} E_*^{\gamma-1/2}\,. 
    \label{eq:exact_vel_drd}
\end{equation}

\begin{figure*}
    \centering
    \includegraphics[width=0.98\linewidth]{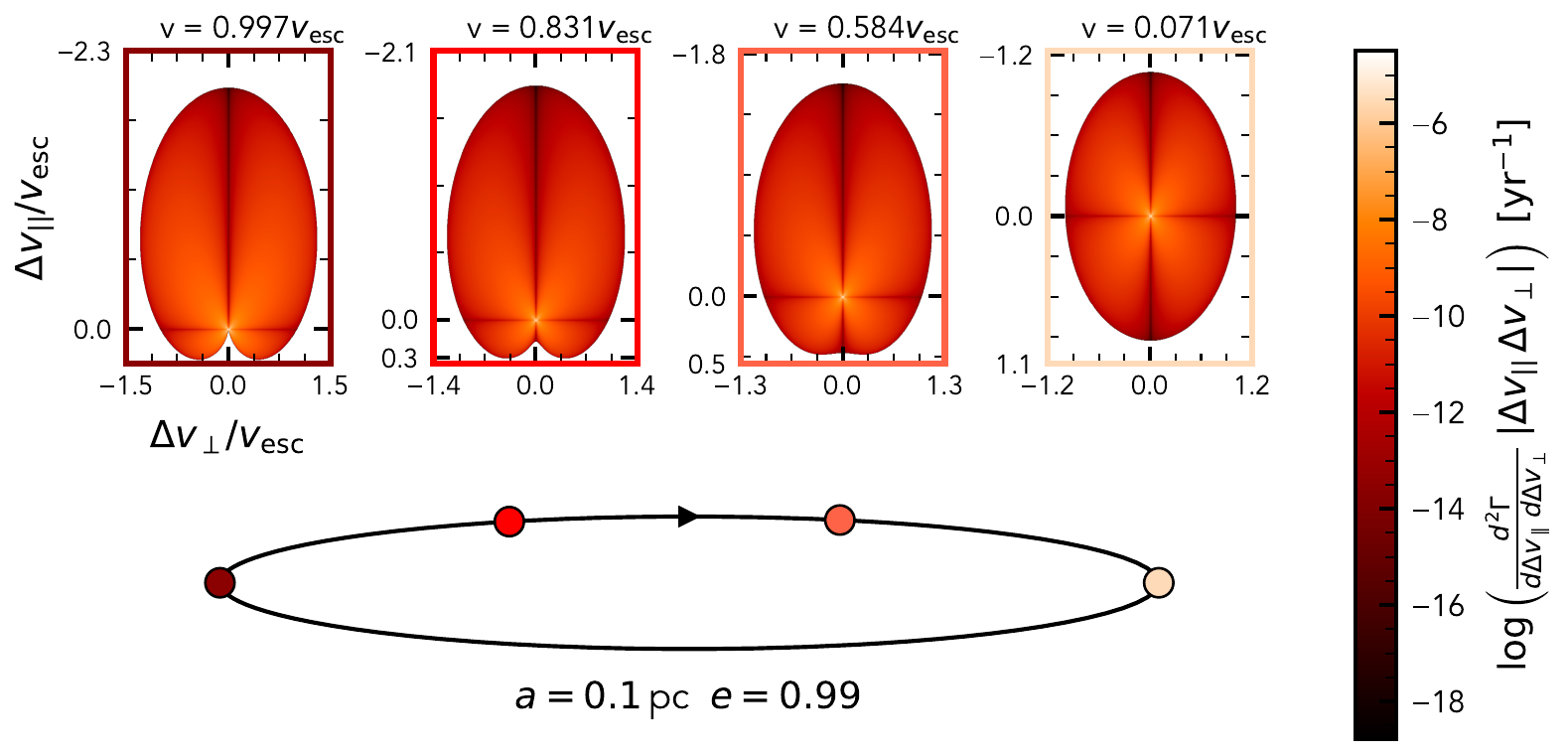}
    \caption{The differential rate distribution, ${\rm d}^2\Gamma/{\rm d}\Delta v_\parallel {\rm d}\Delta v_\perp$, for an object in a Keplerian orbit around an $\Mbh=10^6\,M_\odot$ central object. Here, $\gamma=7/4$, $a=0.1\,\mathrm{pc}$, and the eccentricity $e=0.99$. The scatterers and test object have the same mass $M=M_{\rm s}=1\,M_\odot$. Each panel depicts a different snapshot along the orbit of the test particle, moving from pericenter (left panel) to apocenter (right panel). The colored frames of each panel correspond to the equivalently colored point along the example orbit in the bottom panel. The color shading in each panel corresponds to the log value of  ${\rm d}^2\Gamma/({\rm d}\dvpar {\rm d}\dvper)$, multiplied by $|\dvpar \dvper|$ to have units of rate (yr$^{-1}$). }
    \label{fig:Orbit_kicks}
\end{figure*}
A plot showing the differential rate distribution as a function of $\dvpar$ and $\dvper$ is shown in Figure~\ref{fig:Orbit_kicks}. As the test object traverses its eccentric orbit, the allowed kick distribution changes shape and magnitude. 
Notably, some kick magnitudes are accessible from a single strong scattering only along individual subsections of the orbit; the most extreme kicks are heavily dependent on the mass ratio between the test object and the field object.

Beyond the bivariate DRD in Eq. \ref{eq:exact_vel_drd}, we are also interested in the DRD's marginal distributions, wherein the differential rate can be written as a function of solely $\dvpar$ or $\dvper$ (after integrating over the other velocity kick component). Unfortunately, there is no general closed-form solution for these marginal distributions, even for the simple power-law DFs we consider for NSCs here.  However, it is straightforward to compute these marginal distributions for the special case of $\gamma=3/2$ (where the math is dramatically simplified by the constancy of the DF), and they are
\begin{align}
    \frac{{\rm d}\Gamma}{{\rm d}\dvpar}\Bigg|_{\gamma=\frac{3}{2}} =& 16\pi^2 G^2 M_{\rm s}^2 C_{3/2}\left( \mathcal{F}(\Delta v_{\perp, \max}) - \mathcal{F}(\Delta v_{\perp, \min})\right), \\
    \frac{{\rm d}\Gamma}{{\rm d}\dvper}\Bigg|_{\gamma=\frac{3}{2}} =& 16\pi^2 G^2 M_{\rm s}^2 C_{3/2}\left( \mathcal{G}(\Delta v_{||, \max}) - \mathcal{G}(\Delta v_{||, \min})\right)\,,
\end{align}
with the $\mathcal{F}$ and $\mathcal{G}$ functions written explicitly in Eq.~\ref{eq:curly_F_G}. The minimum and maximum velocity kicks used as inputs for the $\mathcal{F}$ and $\mathcal{G}$ functions come from solving the corresponding extrema from the kick boundary.

To approximate the marginal distribution for arbitrary $\gamma$, 
we begin with the exact marginal distributions above, and then hypothesize that the leading-order difference between scattering rates in varied NSCs will not be the shape of the DF but rather its normalization.  
Since the scattering rate is always proportional to the number density, we now scale the $\gamma=3/2$ result by the ratio $\mathcal{N}(r)$ between the number density for arbitrary $\gamma$ and the number density for $\gamma=3/2$ profiles.  As we shall see, this seemingly crude approach provides a surprisingly accurate approximation, which for the parallel kicks becomes
\begin{equation}
    \frac{{\rm d}\Gamma}{{\rm d}\dvpar} \approx \mathcal{N}(r) \frac{{\rm d}\Gamma}{{\rm d}\dvpar}\Bigg|_{\gamma=\frac{3}{2}} \label{eq:marginal_dvpar_close}\,,
\end{equation}
with the density correction factor $\mathcal{N}(r) = n_*(r) / n_*(r)|_{\gamma=3/2}$. At leading order (i.e. the weak scattering limit)
\begin{equation}
    \frac{{\rm d}\Gamma}{{\rm d}\dvpar} \approx \frac{12 \pi G^2 M_{\rm s}^2 n_* }{v_\mathrm{esc}^3} \big(\frac{G\Mbh}{3r}-\frac{v^2}{10}\big) \frac{1}{|\dvpar|^3}\,.
    \label{eq:weak_marginal_par}
\end{equation}
This approximation is notably similar to the even cruder estimate in Eq.~\ref{eq:twiddle_dist}. Here, we also encounter a similar problem faced by the DCs, namely the ``Coulomb logarithm,'' denoted $\ln(\Lambda) = \ln(b_\mathrm{max}/b_0)$, with $b_\mathrm{max}$ a rough upper bound of the largest viable\footnote{For both the DRD and standard DC theory, ``viable'' here generally refers to the validity of the $n-\Sigma-v$ assumptions used in these types of derivations.  The two assumptions that most typically break down to set $b_{\rm max}$ are (i) uniformity of the background density, i.e. $b_{\rm max} \sim n_\star / ({\rm d} n_\star/{\rm d} r)$, and/or (ii) pairwise interactions, i.e. $b_{\rm max} \sim n_\star^{-1/3}$.} impact parameter. Upon taking the second moment in $\dvpar$ using Eq~\ref{eq:weak_marginal_par}, we see that $\langle \dvpar^2\rangle \propto \ln (\Delta v_{||, \max}/\Delta v_{||, \min})$. We can then identify that $\Delta v_{||,\min} = 2 G M_{\rm s}/(b_\mathrm{max} V)$, recovering the ambiguity of the Coulomb logarithm even in this formalism. 

To recover the classic form of dynamical friction, we notice that the next highest order term in Eq.~\ref{eq:marginal_dvpar_close} scales as $\dvpar^{-2}$. Upon taking the first moment of Eq.~\ref{eq:marginal_dvpar_close}, we identify that the leading-order term containing $\dvpar^{-3}$ vanishes, while the term that scales as $\dvpar^{-2}$ then itself acquires a Coulomb logarithm factor. Thus, we recover basic properties of the DCs through the DRD, where the first and second moments both contain the Coulomb logarithm.

A similar process can be done for the marginal distribution for perpendicular velocity kicks, with there again being a fully analytic solution for $\gamma=3/2$ that can be exploited to approximate a general ${\rm d}\Gamma / {\rm d}\dvper$:
\begin{equation}
    \frac{{\rm d}\Gamma}{{\rm d}\dvper} \approx  \mathcal{N} \frac{{\rm d}\Gamma}{{\rm d}\dvper}\Bigg|_{\gamma=\frac{3}{2}} \label{eq:marginal_per}\,.
\end{equation}
Since the motion of our particle does not introduce headwinds in the $\dvper$ direction, multiplying by $2/3$ for isotropy and substituting $V \sim \sqrt{3} \sigma(r)$ into Eq.~\ref{eq:twiddle_dist} yields
\begin{equation}
    \frac{{\rm d}\Gamma}{{\rm d}\dvper} \approx \frac{16\pi G^2 M_{\rm s}^2 n_*}{3 \sqrt{3} \sigma(r) \dvper^3} 
    \label{eq:weak_marginal_per}\,.
\end{equation}

\begin{figure*}
    \centering
    \includegraphics[width=1\linewidth]{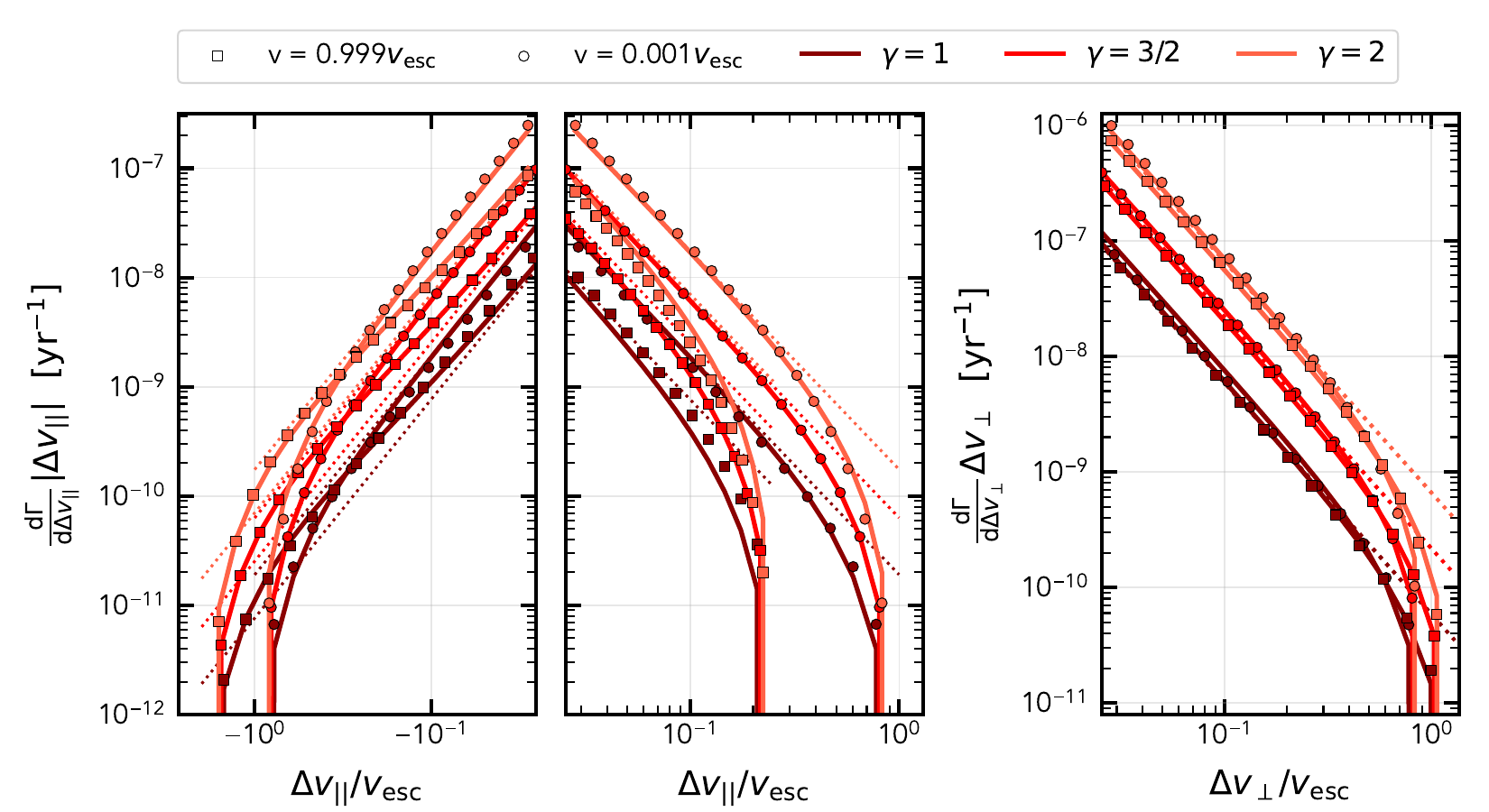}
    \caption{Marginal distributions and approximations thereof for our DRD in a power-law NSC. The marginal distributions and approximations are calculated for
    a set of $\gamma$ values, including $\gamma=1$ (dark red), $\gamma=3/2$ (medium red), and $\gamma=2$ (orange), evaluated at $r=0.1 r_\mathrm{inf}$. 
    The direct integration of Eq.~\ref{eq:exact_vel_drd} is shown for two velocities: $v=0.999v_\mathrm{esc}$ (squares) and $v=0.001v_\mathrm{esc}$ (circles). On the left two panels, the more accurate approximation of Eq.~\ref{eq:marginal_dvpar_close} is shown as a solid line, whereas the cruder Eq.~\ref{eq:weak_marginal_par} approximation ($\propto |\dvpar|^{-3}$) is shown as a thin dotted line. On the rightmost panel, the solid line is Eq.~\ref{eq:marginal_per} and the dotted line is instead Eq.~\ref{eq:weak_marginal_per} ($\propto |\dvper|^{-3}$).  Eqs. \ref{eq:marginal_dvpar_close} and \ref{eq:marginal_per} are  excellent analytic approximations to the full calculation.}
    \label{fig:Marg_dists}
\end{figure*}

Examples of the exact marginal distributions and their approximations are shown in Figure~\ref{fig:Marg_dists}, where we notice that the heuristic scalings yield good results across a broad range of astrophysically plausible $\gamma$ values, from $1$ to $2$. This is because although $\gamma$ changes the velocity distribution, it does so relatively weakly. In contrast, the number density of field stars scales as $n_*\propto r^{-\gamma}$. 

\subsection{Moments for Nuclear Star Clusters} \label{sec:third_mom}
From our previous results, we identify a heavy tail allowed by the strong scatterings in $\dvpar$. Part of this information can be captured in the higher-order moments of the DCs. The moment $\langle \dvpar^i \dvper^j \rangle $  can be calculated as 
\begin{equation}
    \langle \dvpar^i \dvper^j \rangle  = \int \dvpar^i \dvper^j \frac{{\rm d}^2\Gamma}{{\rm d}\dvpar {\rm d} \dvper}\, {\rm d}\dvpar \,{\rm d}\dvper \label{eq:DRD_diffcos} .
\end{equation}
To demonstrate the similarity between this formalism and that of the standard DC calculations (Eq.~\ref{eq:Diff_cos}), we calculated the DCs from both Eq.~\ref{eq:DRD_diffcos} and Eq.~\ref{eq:Diff_cos} letting $v$ range from $5\%$ to $99.9\%$ of $\vesc$, fixing $\gamma=7/4$, $r=0.1\,r_\mathrm{inf}$, $M=M_{\rm s}=1\,M_\odot$, and $\Mbh=10^6\,M_\odot$. We find that the relative error between the DRD-derived and standard formalism has a median $0.8\%$, $1.8\%$, and $1.8\%$ for $\langle \dvpar \rangle$,  $\langle \dvpar^2 \rangle$, and  $\langle \dvper^2 \rangle$, respectively, across this velocity range. The Coulomb-logarithm ambiguity is dealt with by specifying a minimum velocity kick scale for integration to avoid the divergence at 0. We naturally take $\Delta v_\mathrm{min} = 2GM_{\rm s}/(\sqrt{3} \sigma(r) b_\mathrm{max}) $, with $b_\mathrm{max}$ taken to be the same as in the Coulomb logarithm.

Leveraging the same technique of fixing $\gamma=3/2$ to find closed-form expressions for the marginal distributions, the third moment for parallel velocity kicks can be written
\begin{equation}
\begin{gathered}
        \langle \dvpar^3 \rangle \Big|_{\gamma=\frac{3}{2}}  = 
       32 \pi^2 G^2 M_{\rm s}^2 C_{3/2}  \mu \left[- \frac{v v_\mathrm{esc}^2}{5} + \frac{v^3}{21}\right] 
\end{gathered}
\end{equation}

To extend to other values of $\gamma$, the same technique of scaling by the ratio of the number densities provides a sufficient approximation:
\begin{equation}
    \langle \dvpar^3 \rangle \approx  \mathcal{N} \langle \dvpar^3 \rangle \Big|_{\gamma=\frac{3}{2}} \,.
    \label{eq:dvpar3_approx}
\end{equation}


We can repeat this procedure for the other non-zero\footnote{Other third moments of the DRD are zero assuming field star velocity isotropy, though they would attain finite values for most orbits in most anisotropic velocity fields.} third moment in the special case of $\gamma=3/2$:
\begin{equation}
    \langle \dvpar \dvper^2 \rangle \Big|_{\gamma=\frac{3}{2}} = 32\pi^2 G^2 M_{\rm s}^2 C_{3/2} \mu [-  \frac{2 v_\mathrm{esc}^2 v}{15}  + \frac{2}{105} v^3 ]\,.
\end{equation}
And once again, performing a similar scaling procedure, we approximate for general $\gamma$ as:
\begin{equation}
    \langle \dvpar \dvper^2 \rangle \approx \mathcal{N} \langle \dvpar \dvper^2 \rangle \Big|_{\gamma=\frac{3}{2}}  .
    \label{eq:dvpardvper2_approx}
\end{equation}

\begin{figure}
    \centering
    \includegraphics[width=0.98\linewidth]{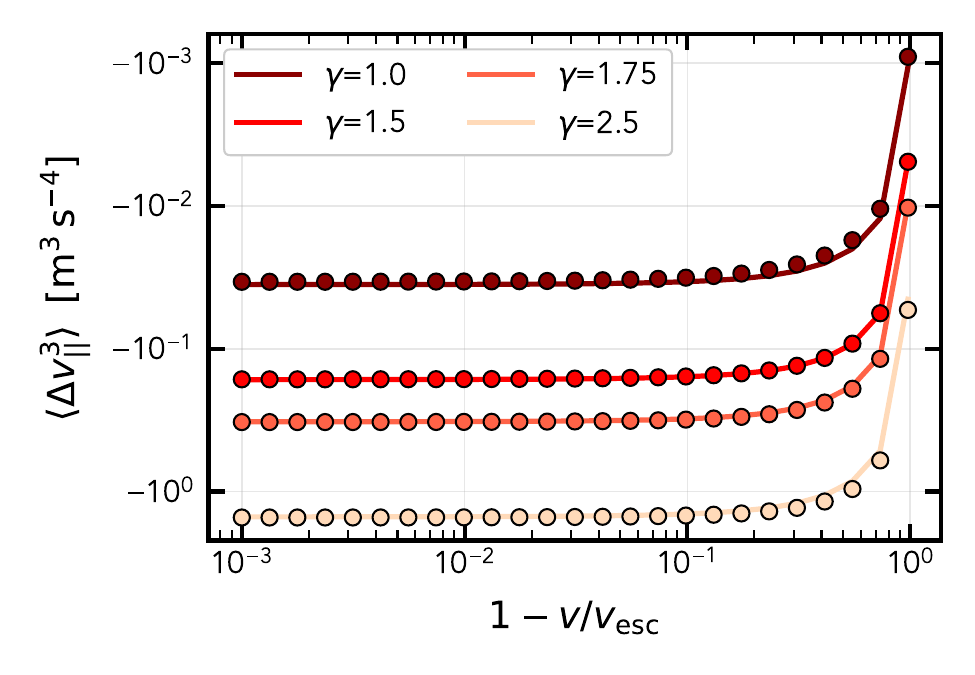}
    \includegraphics[width=0.98\linewidth]{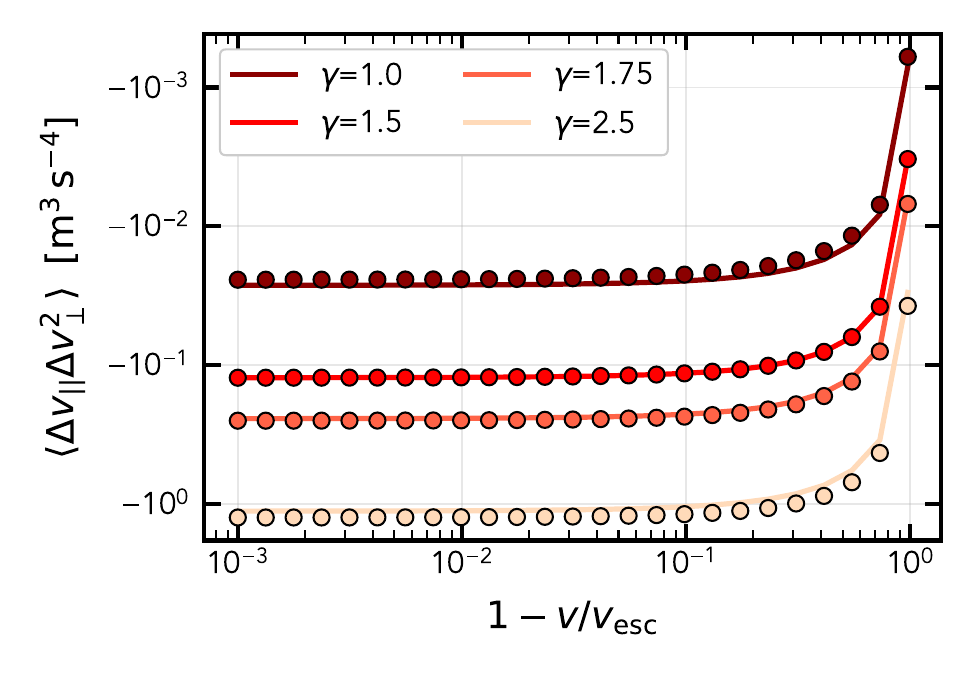}
    \caption{Plots displaying the two nonzero third moments of the DRD, as functions of $1-v/\vesc$. The third moment taken from direct integration of Eq.~\ref{eq:exact_vel_drd} is shown with points of different colors, corresponding to different NSC power-law slopes $\gamma$. We again use $M=M_s=1\,M_\odot$, $M_\bullet = 10^6\,M_\odot$, and $r=0.1 \,r_\mathrm{inf}$. The solid lines in the top panel are from Eq.~\ref{eq:dvpar3_approx}, and in the bottom panel, the solid lines are from Eq.~\ref{eq:dvpardvper2_approx}; in both cases, we see good agreement with exact numerical calculation.}
    \label{fig:third_moment}
\end{figure}

Both of these third moments are presented in Figure~\ref{fig:third_moment}, where we see that the approximations derived above (solid lines) provide an excellent scaling to the numerically calculated exact third moments (circles). The actual values between both third moments are similar:
\begin{equation}
    \frac{\langle \dvpar \dvper^2 \rangle \Big|_{\gamma=\frac{3}{2}}}{\langle \dvpar^3 \rangle \Big|_{\gamma=\frac{3}{2}}} = \frac{-\frac{2}{15} + \frac{2}{105} (\frac{v}{v_\mathrm{esc}})^2}{-\frac{1}{5}+ \frac{1}{21} (\frac{v}{v_\mathrm{esc}})^2}\,.
\end{equation}
For velocities $0 \le v \le v_{\rm esc}$, the ratio of the third moments is bounded
\begin{equation}
    \frac{2}{3} \le \frac{\langle \dvpar \dvper^2 \rangle \Big|_{\gamma=\frac{3}{2}}}{\langle \dvpar^3 \rangle \Big|_{\gamma=\frac{3}{2}}} \le \frac{3}{4}.
\end{equation}

While third moments of the DRD are generally less important for overall rates of energy transport in collisional star systems than are low-order moments \citep{BinneyTremaine}, they can result in anomalous diffusion effects \citep{Bar-or2013}.  More speculatively, by introducing a skewness into the DRD, they may eventually be helpful at regulating numerical issues in studying diffusion in coordinate systems where particular variables have mathematically imposed boundaries, for example diffusion of the Carter constant for Kerr metric geodesics \citep{Alush+26}.

\section{Action Space and Astrophysical Applications} \label{sec:action_applications}
Although working in velocity space has inherent benefits due to the simplicity of the DRD, many types of astrophysical processes instead depend on integrals of motion of the test object, such as energy or AM. 
For an NSC, sufficiently large energy changes dictate ejections from the NSC; likewise, tidal disruption events and extreme mass ratio inspirals are driven (or throttled) by AM evolution.

To express the previously derived rate distribution in terms of the specific energy and AM, it would be ideal to use the Jacobian of the map $(\dvpar, \dvper)\to (\dE, \dJ)$.  In practice, this is substantially complicated by the vector nature of AM. However, in an NSC, the typical orbits relevant for ejections, tidal disruption, and EMRIs are all highly eccentric orbits, with specific angular momentum $J$ being a minute fraction of the circular angular momentum $J_\mathrm{circ}$.  In this section, we will perform a series of approximations appropriate for this highly eccentric limit, deriving simplified but tractable rates of differential action space kicks that can then be straightforwardly applied to problems in NSC astrophysics where strong scatterings may be relevant.


For such an eccentric object with $1-e\ll 1$ in an NSC, there are two regimes of motion: the pericentric regime, comprised of motion that is nearly radial and $v/\vesc\approx 1$; and the apocentric regime, where $v/\vesc\ll 1$, which is where the object spends most of its time. For the radial part of the orbit, perpendicular velocity kicks are nearly orthogonal to the radius vector, and so   
\begin{equation}
\begin{gathered}
    \Delta J \approx  r \dvper \\
   \frac{\dJ^2}{2r^2} +  \dE = - v \dvpar - \dvpar^2/2\,.
\end{gathered}
\end{equation}

Then, the mapping $(\dE, \dJ)\to \dvpar$ is characterized as
\begin{equation}
    \dvpar = -v \pm \sqrt{v^2 - 2\dE - \dJ^2/r^2} \,,
\end{equation}

and thus under this approximation, the Jacobian map is straightforward, 
\begin{equation}
    {\rm d}\dE\, {\rm d}\dJ = r \sqrt{v^2 - 2 \dE - \dJ^2/r^2}\, {\rm d}\dvpar\, {\rm d}\dvper\,.
\end{equation}

Thus, given some $(\dE, \dJ)$, we can map to velocity space and then compute the rate densities for all valid velocity coordinates. Explicitly, the transformation becomes
\begin{equation}
\begin{gathered}
        \frac{{\rm d}^2 \Gamma}{{\rm d} \dE {\rm d} \dJ} =  \sum_i \frac{1}{r \sqrt{v^2 - 2 \dE - \dJ^2/r^2}} \left(\frac{{\rm d}^2\Gamma}{{\rm d} \dvpar {\rm d} \dvper}\right)_i\,. \label{eq:actiondrd}
\end{gathered}
\end{equation}
The subscript $i$ allows for the two different branches of $\dvpar$ allowed by the quadratic nature of the variables. Due to the trivial mapping between $\dvper$ and $\dJ$ under our assumptions, the results from the boundaries of the velocity kick distribution dictate that
\begin{equation}
    \Delta J_\mathrm{max} = r \Delta v_{\perp ,\mathrm{max}}  \approx \frac{3\sqrt{3}}{2}\mu  v_\mathrm{esc} r\,,
\end{equation}
as $\lambda \approx -1/2$ for $v\approx \vesc$. For the energy, a similar approach can be taken to the velocity kick boundary, although the maximum energy kick need not lie along the velocity boundary itself due to the more complex mapping. Instead, expressing $\dvpar(\chi)= \dv \eta \cos(\chi)$ and  $\dvper(\chi)= \dv \eta \sin(\chi)$, where $0 \le \eta \le 1$ allows us to maximize over both $\chi$ and $\eta$. Doing so, we find that the maximum specific energy changes are (with $v\sim \vesc$)

\begin{equation}
    \Delta E_\mathrm{max} = 
    \begin{cases}
        4\mu (1-2\mu) v_\mathrm{esc}^2, \hspace{1cm} 0\le \mu \le 1/4\, \\
        v_\mathrm{esc}^2/2,  \hspace{1cm} 1/4 < \mu <1\,
    \end{cases}
\end{equation}
and that the corresponding minimum energy kicks are
\begin{equation}
    \Delta E_\mathrm{min} = 
    \begin{cases}
        -\frac{\mu}{1-\mu} \frac{v_\mathrm{esc}^2}{2} , \hspace{1cm} 0 \le \mu <3/4 \\
        4\mu (1-2\mu) v_\mathrm{esc}^2, \hspace{1cm} 3/4 \le \mu <1 \,.
    \end{cases}
\end{equation}

\begin{figure*}
    \centering
    \includegraphics[width=0.98\linewidth]{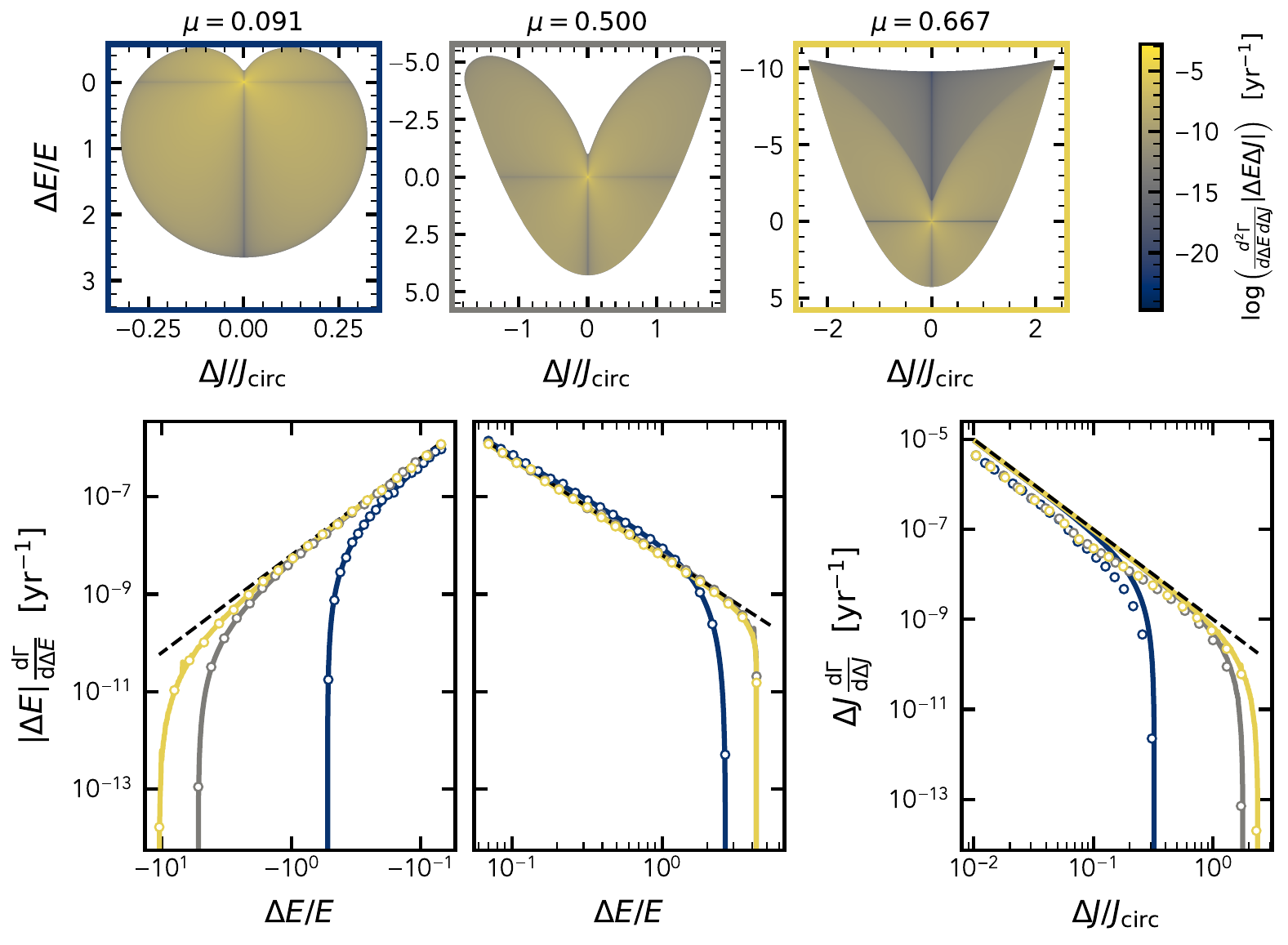}
    \caption{Local 2D and marginal distributions for the DRD of kicks in action space, for an object with orbital parameters $a=0.1 r_\mathrm{inf}$, $e=0.999$, and a location chosen such that $v/\vesc = 0.9$, in its orbit around a $10^6\,M_\odot$ SMBH with $\gamma=7/4$. The top panels display the rate distribution in energy space across varied mass ratio $\mu=M_{\rm s}/(M+M_{\rm s})$ for $M_{\rm s}=1\,M_\odot$ (i.e. corresponding to $M=$ 10, 1, 0.5$\,M_\odot$, from left to right). The colored frame of each plot maps to the corresponding marginal distributions in the bottom panels, where we show changes in the negative energy branch (to unbind the particle, bottom left), the positive energy branch (to bind the particle, bottom middle), and the changes to the AM scaled by the circular AM (bottom right). The points denote more exact numerical estimates, made by direct Monte Carlo sampling of energy and AM kicks that do not make the radial-motion approximation. In general, we see excellent agreement between exact estimates and the radial-motion approximation for ${\rm d}\Gamma/{\rm d}\Delta E$, and reasonable agreement for ${\rm d}\Gamma/{\rm d}\Delta J$. The black dashed lines dictate the weak limit for ${\rm d}\Gamma/{\rm d}\dE$ and ${\rm d}\Gamma/{\rm d}\dJ$. As with $\dvper$ in Fig. \ref{fig:Orbit_kicks}, $\dJ$ is artificially mirrored for viewing in the upper panels.}
    \label{fig:action_marg}
\end{figure*}
The 2D DRD in action space and the corresponding marginal distributions are shown in Figure~\ref{fig:action_marg}, where geometrically complex distributions emerge from the nonlinear Jacobian factor as $\mu$ varies. The specific results shown are for a particle traveling on an orbit with $a=0.1 r_\mathrm{inf}$, with $e=0.999$ and the location chosen such that $v/\vesc = 0.9$ (i.e. a part of the orbit moderately into the ``pericentric'' regime of motion). For large test objects, there is a strong preference for strong encounters with a light field object to more tightly bind them. There is a clear emergence of approximate power law behavior for action space kicks in the marginal distributions, up to the cutoffs set by the aforementioned kick boundaries. This quasi-power-law behavior is expected to be inherited from the velocity space DRD, since the mapping from $\dvper$ to $\dJ$ is simple, and also (for the energy changes at rapid velocity and weak kicks) $\dE \approx -v \dvpar$.

To understand the validity of the approximations we have introduced by assuming a radial orbit, we also show in Fig. \ref{fig:action_marg} data points with the exact marginalized distributions.   These are computed by Monte Carlo sampling of Eq.~\ref{eq:rate_measure}, where we convert the drawn velocity kick variables to changes in energy and AM, with no assumptions made that the orbit is radial. We see that for the marginal distributions in energy, the agreement is excellent. Our approximations work less well for changes in AM.  This can be attributed 
mainly to the fact that our prescription $\dJ \approx r \dvper$ implies
\begin{equation}
    J_f = |\vec{J_f}| = |\vec{J_i} +\vec{\dJ}| \approx J_i + \dJ
\end{equation}
where $J_i$ and $J_f$ are the magnitudes of the AM before and after a scattering, respectively. As the product of the tangential velocity and the radius is conserved through an orbit, the above equation holds best at apocenter, when the tangential velocity is minimized, and vector addition can be approximated by scalar addition in the strong regime. Even in this case, we find that relative errors between the Monte Carlo draws for the marginalized AM and our radial approximation have a median of $ \sim 100\%$, with the error being dominated in the weak regime. For the same parameters but instead $1-e=10^{-4}$, we find the median error drops to $\sim 50\%$, and for $1-e=10^{-5}$, the median relative error falls to be  $\sim 20\%$. These errors are larger than one would ideally hope for, but are still modest enough to enable many astrophysical applications, particularly for the near-radial orbits characteristic of loss cone problems.  We find the energy is well approximated in the radial regime, as we made no underlying approximations about its behavior.
\subsection{Canceled EMRIs}
EMRIs are a key science case for the upcoming Laser Interferometer Space Antenna \citep[{\it LISA};][]{LISA2017}, where they will provide a particularly stringent test of general relativity \citep{Barack2007}. An EMRI is expected to spend many orbits inside the {\it LISA} bandpass before plunging into the SMBH, building up incredible signal-to-noise \citep{Babak2017}. However, a key uncertainty for {\it LISA} is the EMRI rate, as it is unknown whether there will be tens or thousands of EMRIs in the {\it LISA} band at any given moment. We aim below to understand the plausibility of derailing a proto-EMRI from ever reaching the {\it LISA} band, and the corresponding rate at which this ``cancellation'' may occur.

Typically, an EMRI's evolution is governed by two competing regimes: the timescale for GW-inspiral ($t_\mathrm{GW}$) between the test object and the SMBH; and the AM diffusion timescale ($t_\mathrm{AM}$) caused by the sea of field particles around the test object. When $t_\mathrm{GW}\gg t_\mathrm{AM}$, the particle exhibits a random walk in $(1-e, a)$ space, potentially crossing into the GW-dominated regime, becoming circular, or plunging into the SMBH without an appreciable GW-inspiral \citep{2005ApJ...629..362H, Qunbar_stone, Mancieri2025}

If the particle crosses sufficiently into the GW-dominated regime, $t_\mathrm{GW} \ll t_\mathrm{AM}$ and the particle will first circularize from GW at nearly fixed pericenter, and then decrease its semi-major axis until it inspirals into the SMBH \citep{Peters1964}. Late in this process, the particle will enter the {\it LISA} band and provide GW signals that will be key to {\it LISA}'s broader science goals.

In this standard picture, EMRIs totally decouple from stellar perturbations once $t_\mathrm{GW} \ll t_\mathrm{AM}$.  However, one can hypothesize that although AM diffusion appears to be subdominant, there is still a rare chance of a close encounter at apocenter that might send the particle from the GW-dominated regime to the AM-dominated regime, canceling its subsequent inspiral.

Consider a proto-EMRI which is still highly eccentric and has yet to enter the {\it LISA} bandpass, but is nevertheless clearly on the EMRI track ($t_{\rm GW} \ll t_{\rm AM}$). At this stage, $v/\vesc \ll 1$ at apocenter, so the asymmetries in the kick distribution caused by the particle's motion are minute. Thus, we can consider the two-dimensional velocity plane normal to the radius vector at apocenter.  Although kicks in this plane have a geometry which is technically different from the parallel and perpendicular directions we modeled previously, in practice they will behave the same as $\dvper$ (again, due to $v \ll v_{\rm esc}$).  Other kicks out of this plane, i.e. parallel or antiparallel to the radius vector, do not contribute to the AM. In other words, at apocenter
\begin{equation}
    J \approx r v_t , \Delta J \approx r \dvper\,.
\end{equation}

From Eq.~\ref{eq:weak_marginal_per}, the marginal distribution of AM kicks can be expressed as
\begin{equation}
    \frac{{\rm d}\Gamma}{{\rm d}\Delta J} \approx \frac{16\pi G^2 M_{\rm s}^2 n_*}{3 \sqrt{3}\sigma(r) \Delta J^3} r^2\,. \label{eq:CE_twiddle_drd}
\end{equation}




The critical AM $J_c$ at the boundary between the GW and AM-dominated regimes is found by setting  $t_\mathrm{GW}=t_\mathrm{AM}$ \citep{2005ApJ...629..362H},
\begin{equation}
    J_c^{5} = \frac{85}{3} \frac{ k \sigma(a)^3}{n(a) \langle M_{\rm s}^2 \rangle \ln(\Lambda)} \frac{G^{7/2} \Mbh^{9/2} M}{a^{3/2} c^5}\,. \label{eq:am_ce_crit}
\end{equation}
Here, $\langle M_s^2 \rangle$ is the second moment of the mass distribution --- taken to be $(1\,M_\odot)^2$ --- and $k\approx 0.34$ \citep{BinneyTremaine}.
The cancellation of a proto-EMRI requires an encounter to deposit $\Delta J \approx J_c$ to the object to move it to the relaxation-dominated regime. The instantaneous canceled EMRI rate $\Gamma_{CE}$ can be written using Eq.~\ref{eq:CE_twiddle_drd}:
 \begin{equation}\begin{gathered}
          \Gamma_\mathrm{CE} = \int_{\Delta v_{\perp,c}}^{\Delta v_{\perp, \max} } \frac{{\rm d}\Gamma}{ {\rm d} \dvper} {\rm d} \dvper \\ \\ \approx  \frac{8\pi G^2 M_{\rm s}^2 n_*(r) }{3 \sqrt{3} \sigma(r)} [\frac{1}{\Delta v_c^2} - \frac{1}{\Delta v_{\perp, \max}^2 }]\,.
 \end{gathered}
 \end{equation}
 Here, $\Delta v_{c} = (J_c-J)/r$, and $\Delta v_{\perp, \max}$ is known from the kick boundary. Since $\dJ \propto r$, cancellation will occur primarily at apocenter. Since the object spends the vast majority of its time near apocenter for an eccentric orbit, the orbit-averaged rate of the object is just its rate calculated at apocenter.

Here, since $v\sim 0$, $\Delta J_\mathrm{max} = 2\mu r_a v_\mathrm{esc}(r_a)$, and so then our orbit-averaged apocentric cancellation rate is 
\begin{equation}
\begin{gathered}
        \Gamma_{\mathrm{CE},o} \approx \frac{8\pi G^2 M_{\rm s} \rho_\mathrm{inf} r_\mathrm{inf}^\gamma  r_a^{2-\gamma} }{3 \sqrt{3}\sigma(r_a) } \left[\frac{1}{(J_c-J)^2} - \frac{1}{8\mu^2 r_a G \Mbh} \right]
\end{gathered}
\end{equation}

\begin{figure}
    \centering
    \includegraphics[width=0.95\linewidth]{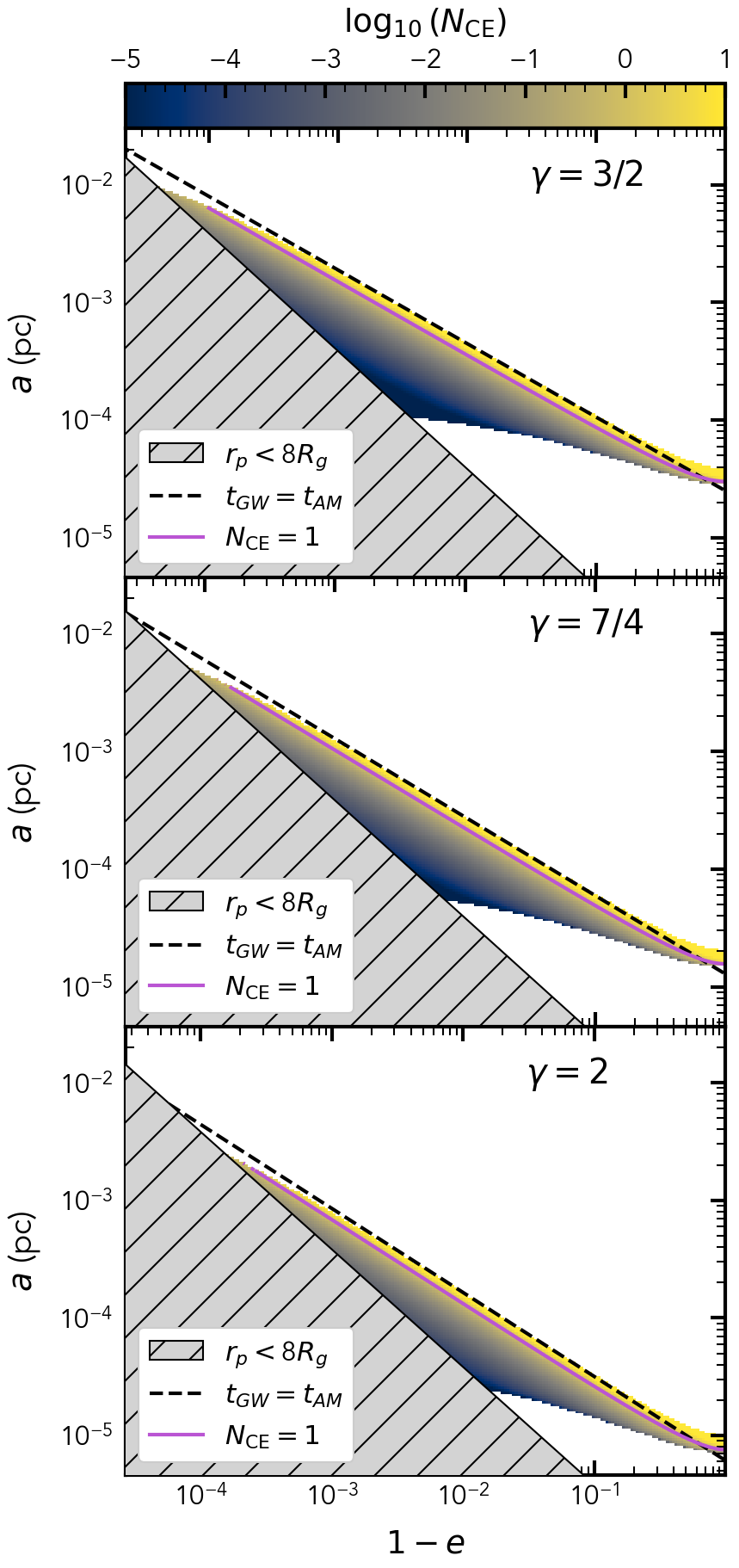}
    \caption{Accumulated numbers of EMRI cancellations, $N_{\rm CE}$, in the space of semimajor axis $a$ and eccentricity $e$, within the GW-dominated zone (i.e. where $t_{\rm GW}<t_{\rm AM}$).   $\Mbh$, $M$, and $M_{\rm s}$ are fixed at $10^6$, $10$, and $1\,M_\odot$, respectively. The top panel shows $\gamma=3/2$, the middle displays the Bahcall-Wolf solution of $\gamma=7/4$, and the bottom panel shows $\gamma=2$. The magenta contour in each plot displays where $N_\mathrm{CE}=1$. We exclude points where $J_c-J < 5\sqrt{\langle \dJ^2 \rangle_o P}$.  We see that for all $\gamma$ values, the $N_{\rm CE}$ contour lies below the $t_{\rm GW}=t_{\rm AM}$ contour, indicating that EMRI cancellation will suppress the low-eccentricity tail of observed EMRI distributions.}
    \label{fig:EMRI_rates_tot}
\end{figure}

Notably, all particles that begin on a track of $t_\mathrm{GW}= p t_\mathrm{AM}$, with $p$ being a dimensionless constant less than one, have the same expected number of cancellation encounters. To understand this relation, we first notice that the AM of the object is approximately constant during the initial stages of GW inspiral, and so the pericentric distance is also approximately constant. From Eq.~\ref{eq:am_ce_crit}, $J_c\propto a^\frac{(\gamma-3)}{5}$. Then, we observe that
\begin{equation}
    \Gamma_{\mathrm{CE},o} \propto \frac{a^{5/2-\gamma}}{J_c^2} \propto a^{\frac{5}{2}-\gamma } a^{-\frac{2(\gamma-3)}{5}}.
\end{equation}
Since the AM or pericenter is nearly constant along the high-$e$ phase of the inspiral, we can write that
\begin{equation}
    r_{p} \propto J^2 \propto J_c^2 \propto a^\frac{2(\gamma-3)}{5}.
\end{equation}
We have from \cite{Peters1964} that the GW timescale scales as
\begin{equation}
    t_\mathrm{GW} \propto a^{4} (1-e^2)^{7/2} \propto a^{\frac{1}{2}+ \frac{7(\gamma-3)}{5}},
\end{equation}
and thus the total number is
\begin{equation}
    N_{\mathrm{CE}} \sim \Gamma_{\mathrm{CE},o} t_\mathrm{GW} \propto a^0
\end{equation}
Thus, for a given set of objects that all start along the same line of constant $t_\mathrm{GW} = p t_\mathrm{AM}$, we have an effectively constant canceled EMRI number. It is the choice of initial AM (and thus pericenter) that allows for such nice cancellation. 

The calculation of $N_\mathrm{CE}$ across the relevant $(1-e,a)$ space is shown in Figure~\ref{fig:EMRI_rates_tot}, where we see a large disparity in the expected number of accumulated strong kicks, depending on the location of the start of the GW inspiral. To prevent any potential cancellation being swamped by diffusion of AM, we exclude points where $J_c-J < 5\sqrt{\langle \dJ^2 \rangle_o P}$, or where the change in AM needed from one strong encounter is less than five typical diffusion step sizes. Notably, tracks even well within $t_\mathrm{GW}=t_\mathrm{AM}$ are subject to finite probabilities of strong encounters during their lifetime, although the $N_{\rm CE}=1$ curves are quite close to the $t_{\rm GW}=t_{\rm AM}$ one. 

In this calculation, we did not consider the resulting changes in {\it energy} from velocity kicks, which would result in a vertical displacement in $(1-e, a)$ space. However, to a first approximation, a kick at apocenter will likely not change the energy significantly. Consider an object undergoing a scattering at apocenter on an eccentric orbit, where after scattering, the final velocity is a factor $g$ times the initial velocity ($v_F=g v_I$). Then, the ratios of the initial to final energy can be written
\begin{equation}
    \frac{E_I}{E_F} = \frac{1-v_I^2/\vesc^2}{1-g^2 v_I^2 / \vesc^2} \approx 1+ (g^2-1) \frac{v_I^2}{\vesc^2}\,,
\end{equation}
with the final term truncated to first order. From the vis viva equation at apocenter,
\begin{equation}
    \frac{E_I}{E_F} \approx 1+(g^2-1) \frac{1-e_I}{2}\,.
\end{equation}
Thus, for an orbit with $1-e_I \sim 10^{-2}-10^{-3}$, $g\lesssim 10^{1}-10^{2}$ to stay within an order-unity correction. Thus, if some projection of this velocity kick is in the plane normal to the radius vector, then the AM can change by nearly an order of magnitude while marginally changing the energy.

The net effect of cancellations may not change overall EMRI rates in steady state, where the cancellation phenomenon can be viewed (averaging over many cycles of cancellation and relaxation) as a slowdown in the rate of AM diffusion towards the $t_{\rm GW}=t_{\rm AM}$ separatrix.  However, a potential consequence of EMRI cancellations is a shift in the distribution of eccentricities observed by \textit{LISA}. As an object undergoes the GW-dominated evolution, it will cross the regime where $N_\mathrm{CE}\sim 1$, at which point it will either: (i) be moved back to the AM-dominated regime, potentially wandering back near the GW-dominated regime; or (ii) be placed onto an even more eccentric orbit, where $N_\mathrm{CE}\ll 1$, and it can safely continue on the GW-inspiral track. Thus, strong scatterings will likely increase the number of eccentric EMRIs entering the \textit{LISA} band, while suppressing the number of low-$e$ EMRIs, although direct/MC simulations would be needed to verify this hypothesis.

\subsection{Ejection Rates}
The rates of interesting phenomena like TDEs and EMRIs are governed by the distribution of objects near the SMBH and their diffusion into low-AM orbits vulnerable to tidal disruption or GW-driven inspiral. 
However, processes like ejections can act to throttle the supply of particles into this regime, a concept more generally referred to as ``loss cone shielding'' \citep{Teboul2024}. Ejections act as a leak in the TDE/EMRI supply line, and precise estimates of ejection rates are therefore needed to accurately predict the corresponding rates of loss cone phenomena.

A key advantage of the action space coordinates is their ability to predict outcomes of strong scatterings, such as ejections, which is straightforward once given the marginal distribution $\frac{d\Gamma}{d\dE}$. During the radial part of the eccentric orbit, the object moves at $v\approx \vesc$ and only needs a slight velocity kick to become unbound from the SMBH. To approximate the rate of ejections, one can utilize the leading order terms in Eq.~\ref{eq:weak_marginal_par}; taking $\dE \approx  -v \dvpar$,
\begin{equation}
    \frac{{\rm d}\Gamma}{{\rm d}\dE} \approx \frac{12 \pi G^2 M_{\rm s}^2 n_* v^2}{v_\mathrm{esc}^3} \left(\frac{G\Mbh}{3r}-\frac{v^2}{10}\right) \frac{1}{|\dE|^3}\,. \label{eq:weak_energy_marg}
\end{equation}
An object becomes unbound when $\dE+E<0$, and so the instantaneous local ejection rate is
\begin{equation}
    \Gamma_\mathrm{ej} =   \int_{\Delta E_\mathrm{min}}^{-E} \frac{{\rm d}\Gamma}{{\rm d}\dE}\, {\rm d}\dE \, ,
\end{equation}
where $\Delta E_{\min}$ is the local maximum allowed (negative) energy kick. Using Eq~\ref{eq:weak_energy_marg} and under the assumption of $E\ll |\Delta E_\mathrm{min}|$ along with $v\approx \vesc$, the ejection rate becomes
\begin{equation}
        \Gamma_\mathrm{ej} \approx  \frac{2^{3-\gamma}}{5} \pi  a^2 \rho_\mathrm{inf} v_\mathrm{esc}^{2\gamma+1} \frac{M_{\rm s}}{ \Mbh^2} \left(\frac{G\Mbh}{r_\mathrm{inf}}\right)^{-\gamma} \,.
\end{equation}
A more useful expression is the orbit-averaged ejection rate. Using Eq.~\ref{eq:orb_avg} in the pericentric regime, we have 
\begin{equation}
     \langle \Gamma_\mathrm{ej} \rangle \approx \frac{16\pi}{5P(\gamma-1)} \frac{M_{\rm s}}{\Mbh^2} r_\mathrm{inf}^{\gamma} \rho_\mathrm{inf} a^{3-\gamma} (1-e)^{1-\gamma}
     \label{eq:gamma_ej_twiddle_oa}\,.
\end{equation}

\begin{figure}
    \centering
    \includegraphics[width=0.98\linewidth]{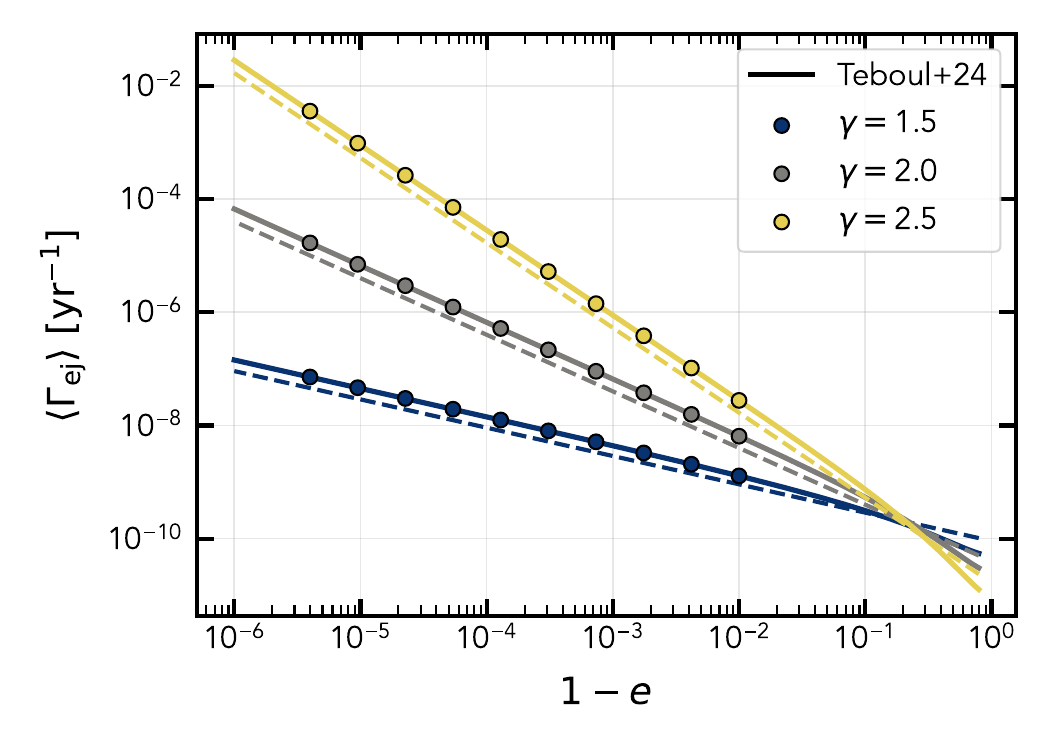}
    \includegraphics[width=0.98\linewidth]{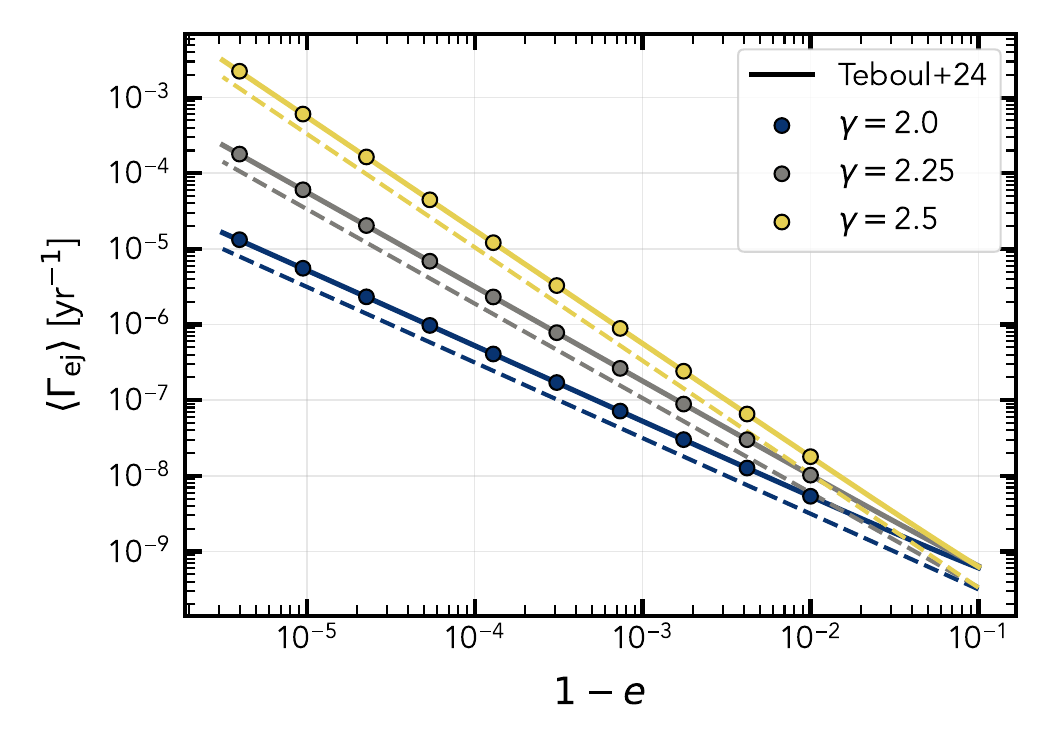}
    \caption{Ejection rates calculated directly from the marginalized DRD in action space (circles), compared against the methodology from \citet[][solid lines]{Teboul2024}, and the twiddle level rates from Eq.~\ref{eq:gamma_ej_twiddle_oa} (dashed lines). Here the central mass is $10^6\,M_\odot$, $M=1\,M_\odot$, and the semi-major axis of the orbit is $r_\mathrm{inf}$. Top panel: Ejection rates for equal mass scatterers $M_{\rm s}=1\,M_\odot$ for $\gamma = $ 1.5, 2, and 2.5. Bottom panel: the scatterers are $M_{\rm s} = 30\,M_\odot$ stellar mass BHs with density profile slopes $\gamma = $ 2, 2.25, 2.5; here we fixed the number $N_\mathrm{BH}=10^3$ inside $r_\mathrm{inf}$. In this case, $r_\mathrm{inf}$ is set by a background stellar density profile of $\gamma=7/4$.}
    \label{fig:Ejection_rates}
\end{figure}

We display our calculations of the orbit-averaged ejection rates in Figure~\ref{fig:Ejection_rates}. We see excellent agreement comparing the ejection rates from Eq.~\ref{eq:actiondrd} and the rates calculated from \cite{Teboul2024}, along with a close approximation from Eq.~\ref{eq:gamma_ej_twiddle_oa}.

However, a key assumption in both the previous ejection rate and the calculation from \cite{Teboul2024} is that kicks capable of ejecting a test star are rare, such that they will likely not happen more than once per orbit. To investigate when this assumption breaks down, we first define the ``saturated strong kick regime''
\begin{equation}
    \frac{{\rm d}\Gamma}{{\rm d}\dE} |\dE| t_c \ge 1\,,
\end{equation}
with $t_c$ being the characteristic timescale over which our actual DRD changes. When this equation is satisfied for $-\dE >E$, there is at least one strong kick that can act to eject the particle. The characteristic radius at which the DRD changes is
\begin{equation}
    r_c = \frac{\frac{{\rm d}\Gamma}{{\rm d}\dE} (r)}{|\frac{{\rm d}\Gamma}{{\rm d}\Delta  E}'(r) |}  \approx \frac{r}{\gamma+1/2}\,
\end{equation}
where we approximated $v\approx v_\mathrm{esc}$ and have taken the weak limit using Eq.~\ref{eq:weak_energy_marg}. The characteristic timescale is simply the crossing time of $r_c$
\begin{equation}
    t_c = \frac{r_c}{v_\mathrm{esc}} = \frac{r^{3/2}}{(\gamma+1/2) \sqrt{2 G \Mbh}}\,.
\end{equation}
We can then solve for $\Delta E_c$, the critical energy for which we expect to have one encounter during the time $t_c$:
\begin{equation}
    \Delta E_c = \sqrt{\frac{4\pi}{5(\gamma+1/2)}} G M_{\rm s}^{1/2} \rho_\mathrm{inf}^{1/2} r_\mathrm{inf}^{\gamma/2} r^\frac{1-\gamma}{2}
\end{equation}
The value of $\Delta E_c$ will fluctuate throughout the orbit, but $E$ will not. The scaling implied by the above equation shows that for $\gamma>1$, the maximum value of $\Delta E_c$ is at pericenter, and so enforcing $E=\Delta E_{c, \max}$ yields the characteristic semimajor axis beyond which we are in the saturated ejection regime:
\begin{equation}
    a_{\rm sat}^\frac{\gamma-3}{2} = \sqrt{\frac{16 \pi}{5 (\gamma+1/2)}} \frac{M_{\rm s}^{1/2}}{\Mbh} \rho_\mathrm{inf}^{1/2} r_\mathrm{inf}^{\gamma/2} (1-e)^\frac{1-\gamma}{2}\,.
\end{equation}
Another assumption implicit in this analysis is that diffusion in energy from weak scatterings alone is much slower than the orbital period, namely that 
\begin{equation}
    t_E \equiv \frac{E^2}{\langle \dE^2 \rangle_o} \gg P,
\end{equation}
where $\langle \dE^2 \rangle_o$ is an orbit-averaged energy diffusion coefficient encapsulating the effects of weak scatterings. 
If energy relaxation happens at or faster than the orbital timescale, strong kicks that might have led to an ejection in isolation can ultimately be swamped by the rapid diffusion from weak encounters, leading to potentially enhanced loss cone shielding.

During the radial part of the orbit, where most of the diffusion in energy happens, we can take $v\approx \vesc$; then the local second-order DC in energy is approximated by (Appendix~\ref{sec:appendix_dc})
\begin{equation}
\begin{gathered}
    \langle \dE^2 \rangle
    \approx \frac{ 64}{3} \pi^2 G^2 M_{\rm s}^2 \ln(\Lambda) C_\gamma \bigg(\frac{G\Mbh}{r} \bigg)^{\gamma + 1/2} B_\gamma \label{eq:dE2_loc_approx}\,.
    \end{gathered}
\end{equation}
Here, $B_\gamma= B(5/2, \gamma-1/2)$, with $B$ being the beta function.
We take $\ln(\Lambda) = \ln (0.4 \Mbh/M_{\rm s})$. Performing an approximate orbit-average with $v_r\approx \vesc$,
\begin{equation}
\begin{gathered}
    \langle \dE ^2 \rangle_o
    = \frac{128 \pi^2 G^2 M_{\rm s}^2 \ln(\Lambda) C_\gamma B_\gamma}{3\sqrt{2}P} \frac{(G\Mbh)^{\gamma}}{(1-\gamma)}\\
    \times a^{1-\gamma} \bigg( 2^{1-\gamma}-(1-e)^{1-\gamma}\bigg)\,.
\end{gathered}
\end{equation}

To find situations where diffusion can swamp singular ejections ($t_E \lesssim P$), setting $t_E = P$ yields the minimum semimajor axis for the ``diffusively saturated regime,''
\begin{equation}
    a_{\rm diff}^{3-\gamma} = \frac{3 \sqrt{2} (G \Mbh)^{2-\gamma} (1-\gamma)}{512 \pi^2 M_{\rm s}^2 G^2 \ln(\Lambda) C_\gamma B_\gamma} \left( 2^{1-\gamma}-(1-e)^{1-\gamma}\right)^{-1}\,.
\end{equation}

Because ${\rm d}\Gamma/{\rm d}\Delta E \propto |\Delta E|^{-3}$, strong scatterings contribute a fraction $\sim 1/\ln\Lambda$ to the total time-averaged rate of energy evolution.  Outside of the diffusively saturated regime, it is {\it only} rare strong scatterings that can eject stars from the system, but in this more exotic regime, even weak scatterings may in principle trigger an ejection over the course of a single orbit, and the greater number of such weak kicks means that they will (for $a > a_{\rm diff}$) dominate ejection rates.  

\begin{figure}
    \centering
    \includegraphics[width=0.95\linewidth]{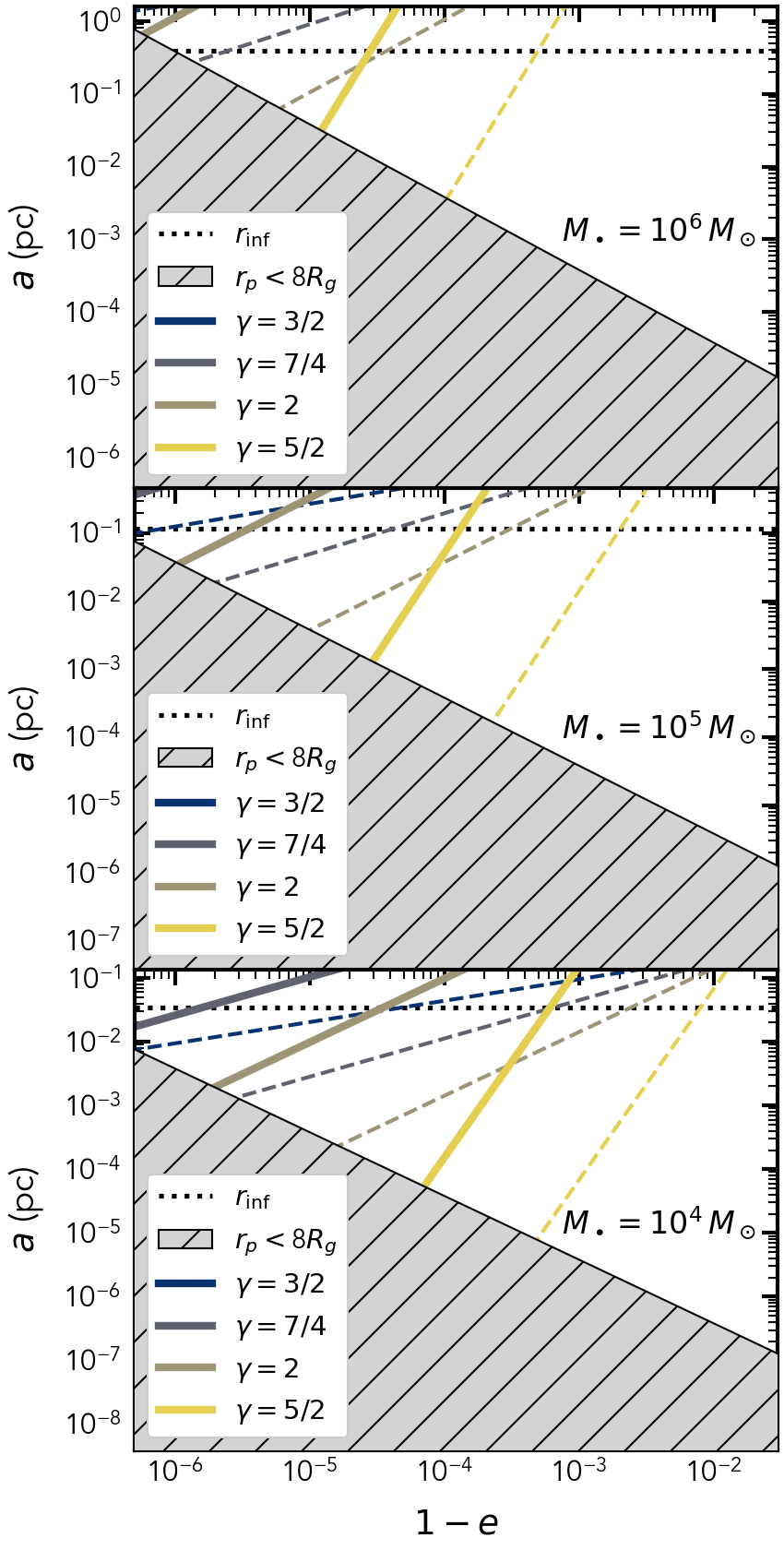}

    \caption{Plots showing different ejection regimes in the space of semimajor axis $a$ and eccentricity $e$, for black hole masses of $10^6\,M_\odot$ (top panel), $10^5\,M_\odot$ (middle panel), and $10^4\,M_\odot$ (bottom panel). All systems take $M=M_{\rm s}=1\,M_\odot$. Solid lines show $\Delta E_{c,\max}=E$ for different values of $\gamma$ (color-coded as per the inset key).  Regions to the left of these solid lines are in the saturated kick regime. The dashed lines indicate where $t_E=P$ for each system of the same color; regions left of these are in the diffusively saturated kick regime.  Saturation of ejections via strong scatterings becomes important for small $M_\bullet$, large $a$, and large $\gamma$. The hatched region is where $r_p \le 8 R_g$, with $R_g=GM_\bullet/c^2$, a typical minimum pericenter to avoid a plunge. The dotted line notates the $r_\mathrm{inf}$ corresponding to $\gamma=7/4$.}
    \label{fig:a_e_ejections}
\end{figure}

Figure~\ref{fig:a_e_ejections} displays both the saturated ejection regime, where $\Delta E_{c,\max}=E$ (solid lines), and the diffusively saturated regime, where $\langle \Delta E^2 \rangle_o P = E^2$ (dashed lines). 
In these regimes, the single kick ejection approximation criterion likely misrepresents the true ejection rate, and regions possessing one or both of the aforementioned criteria become especially significant for a smaller-mass SMBH.  We see in Figure \ref{fig:a_e_ejections} that the diffusively saturated regime is always broader than the saturated strong kick regime, implying that saturation of diffusive energy evolution will occur first as test stars random walk to extreme values of eccentricity.  

The exact implications of energy kick saturation (in either form) for loss cone shielding are not trivial, and we defer a detailed examination of this for future work.  In the saturated regime, test stars can move very quickly to either (i) lower $|E|$, enabling ejections, or to (ii) higher $|E|$, which may eventually freeze out rapid energy evolution (from both strong and weak scatterings).  The first of these processes is itself subtle, as NSCs are embedded in a broader galactic potential, so that even if a test star is kicked to $v > v_{\rm esc} = \sqrt{2GM_\bullet/r}$, it is unlikely to be unbound from the galaxy itself; however, this crude ejection criterion may suffice in practical terms if the stellar potential outside the NSC is sufficiently aspherical \citep{Stone2012}.  The net result of the competition between these two processes is not obvious, and in principle could either greatly strengthen loss cone shielding, if process (i) predominates, or weaken it, if process (ii) does.  Notably, however, Figure \ref{fig:a_e_ejections} shows that saturation of energy kicks occurs in precisely those parts of parameter space (large $\gamma$, large $a$) where loss cone shielding through rare strong scatterings is itself most effective \citep{Teboul2024, TeboulPerets2025}.

Saturation of energy-space evolution also calls into question the accuracy of orbit-averaged schemes to reliably estimate other loss-cone processes around a low-mass SMBH. ``Cliffhanger'' EMRIs are EMRIs that originate in the $t_\mathrm{GW} \gg t_\mathrm{AM}$ regime, far above the canonical ``critical semimajor axis'' that separates true EMRIs from failed plunges \citep{2005ApJ...629..362H}, but which eventually attain $t_\mathrm{GW}\lesssim P$.  They are thus able to dissipate such a significant amount of energy in one orbit that they may move non-diffusively into the $t_\mathrm{GW} \lesssim t_\mathrm{AM}$ regime \citep{Qunbar_stone}, hanging on and becoming true EMRIs. Interestingly, there appears to be a discrepancy in the rates of cliffhanger EMRI success between orbit-averaged and local Monte Carlo schemes for solving the Langevin equation, especially at lower SMBH masses \citep{Mancieri2025}.  The origin of this discrepancy, in which orbit-averaging reduces the cliffhanger rate (i.e. increases the plunge rate), remains unclear, but we speculate\footnote{This speculation is testable by comparing direct and orbit-averaged Monte Carlo schemes in shallow cusps with $\gamma < 3/2$, as this will almost always eliminate the regime of diffusive kick saturation (Figure \ref{fig:a_e_ejections}).} that this could be attributed to the incredibly rapid relaxation in energy that may often coincide with the cliffhanger regime, see Figure~\ref{fig:a_e_ejections}. As ejected objects cannot contribute to the EMRI or plunge rate, they are usually neglected in EMRI statistics for Langevin-type simulations \citep{AlexanderHopman2009, Qunbar_stone, Mancieri2025}, and so regardless of the net direction of energy evolution, the submersion of the $t_{\rm GW} < P$ zone inside the diffusively\footnote{Evidence for this discrepancy comes from the detailed local simulations of \citet{Mancieri2025}.  Because these local Monte Carlo simulations perturb test orbits via diffusion coefficients, strong scatterings are excluded by construction, and thus it is the diffusively saturated dynamics alone that could be relevant.} saturated kick regime is likely to bias EMRI statistics towards fewer plunges, and more apparent cliffhangers.  

Overall, obtaining accurate ejection rates in the regimes where $ \langle \dE^2 \rangle_o P \gtrsim E^2$ or where $\Delta E_{c,\max}\gtrsim E$ using an approach that only considers rates of singular kicks is unlikely to be accurate. Instead, N-body simulations, local Langevin equation integrations, or schemes that directly sampling the full DRD are likely the best way to obtain the ejection rate when energy diffusion is rapid or when the ejection kick regime is saturated.

\section{Conclusion}\label{sec:conclusion}
In this work, we have derived the explicit differential rate distribution in velocity space for the two prototypical examples of collisional star systems: globular and nuclear star clusters.  We have also computed the explicit boundaries of this DRD. For the NSC, we determined approximate formulae for the marginal distributions of kicks and also estimated the third moments of the overall kick rate distribution. We then recasted the velocity kicks into action space using a series of approximations appropriate for loss cone problems in NSCs, and applied this framework to study loss cone shielding via ejections, and the novel phenomenon of EMRI cancellation.

Past work on the collisional dynamics of dense star systems has almost always operated in the limit of weak encounters and diffusive evolution \citep{Spitzer1971, BinneyTremaine}.  The motivations for this choice are clear enough: weak encounters dominate overall rates of energy evolution by a factor of $\sim \ln \Lambda$.  However, even if uncommon strong scatterings are subdominant drivers of the bulk evolution of their host cluster, they may unlock qualitatively novel phenomena forbidden to even a lengthy sequence of weak, diffusive encounters.  The classic example of this kind of non-diffusive effect is stellar ejection \citep{LinTremaine1980}, which has recently seen a revival of interest as a mechanism for shielding SMBH loss cones \citep{Teboul2024} and suppressing rates of transients such as TDEs \citep{TeboulPerets2025} and EMRIs \citep{Kaur2024}.  Other non-diffusive types of evolution, such as the cliffhanger EMRI phenomenon, have also been the topic of recent interest \citep{Qunbar_stone, Mancieri2025}.

Past efforts to systematically study the effects of strong scatterings have usually operated either with transition probability operators \citep{Goodman1983, Bar-or2013} or by making precise rate estimates for a single process enabled by strong scatterings, such as ejection \citep{Teboul2024}.  While transition probability operators are a completely general formalism for handling non-diffusive evolution, their mathematical complexity has so far restricted their application to 1D problems in dense star cluster dynamics (i.e. energy-only evolution).  Our calculation of an explicit differential rate distribution encompasses the exact same physics as a transition probability operator, but is more analytically tractable by working entirely in the differential (rather than integral) mode.  As a result, we are able to compute exact bivariate DRDs in velocity space; while we did not find closed-form translations of the DRD into action space, we present useful approximations, and note that this translation is straightforward to accomplish numerically.

For ejections of high-$e$ stars, we can use the DRD to recover the same orbit-averaged ejection rates as \cite{Teboul2024}, but we also identify new regimes where the assumption of strong scatterings being rare breaks down. First, we identified a regime where strong scatterings can happen more than once per orbit, likely leading to a Poisson-like regime for determining ejection rates. Second, we found that this regime, in which strong kicks saturate, is likely encompassed and superseded by a wider regime in which even weak scatterings lead to non-diffusive evolution of energy.  This broader regime of diffusive energy-kick saturation (where $t_E< P$) may feature surprising implications for loss cone feeding rates, by either strengthening or attenuating loss cone shielding.  We did not quantify these implications in this work, but we did map out the parameter space where they are likely relevant, and found that diffusive kick saturation applies to much of the parameter space ($a\sim r_{\rm inf}$, $\gamma \gtrsim 2$) where standard loss cone shielding has been found to be important \citep{Teboul2024}.  

In applying the DRD formalism to EMRIs, we found that a sizable fraction of EMRIs with $t_\mathrm{GW}<t_\mathrm{AM}$ can undergo a scattering that sends them back into the AM-dominated regime.  The cancellation of these EMRIs will preferentially select against the most circular EMRIs, biasing the loss cone channel for EMRI formation even further towards high-eccentricity signals, and increasing its distinguishability against other formation scenarios, such as Hills separations \citep{Miller2005, Sersante2026} and wet EMRIs \citep{Levin2007}.

It is important to note the assumptions inherent in our analyses. The largest assumption was the choice of a simplistic DF, as most of this paper assumed a DF that was single-mass, spherically symmetric, and isotropic, none of which fully describe the Milky Way's NSC \citep{Fritz2016, Pinna2021}. In principle, our formalism can be generalized straightforwardly to multi-mass or anisotropic systems; asphericity would introduce greater complications \citep{Vasiliev2013}.  During the scattering process, we have also omitted tidal \citep{Lee1986} and relativistic \citep{Oleary2009} effects, which can act to dissipate energy throughout particularly close encounters. Regarding the canceled EMRI number calculation, a more correct approach would be to fully integrate the rate throughout the expected GW-inspiral track \citep{Peters1964}, potentially even using an MC approach to include both DC and strong scattering effects. 

There are likely many other aspects of collisional star clusters that could be studied with the DRD derived in this work.  Particularly likely applications include rate calculations for 
interesting dynamical encounters, including the production of hypervelocity stars in pairwise scatterings \citep{Oleary2008}, close encounters producing LIGO-band gravitational waves through capture of compact objects \citep{Oleary2009}, or depletion of dark matter particles in the center of galaxies \citep{Sharpe2026}.  While we have examined some implications of the DRD for TDEs and EMRIs, others likely exist as well, such as occasional refill of even deeply empty loss cones \citep{Weissbein2017}, which carries implications for both the formation of repeating partial TDEs \citep{Broggi2024} and the eccentricities of EMRIs.

\begin{acknowledgments}
AD and NCS are happy to acknowledge helpful discussions with Yael Alush during early stages of the project. AD gratefully acknowledges support from the \textit{Peter Livingston Scholars Program} at the University of Wisconsin–Madison, made possible by the generosity of Sharon Stark in memory of her husband Peter Livingston. AD acknowledges support from the Fannie and John Hertz Foundation.  Support for NSC and AD was provided by the University of Wisconsin–Madison, Office of the Vice Chancellor for Research and Graduate Education, with funding from the Wisconsin Alumni Research Foundation.  
This material is based upon work supported
by the National Science Foundation Graduate Research Fellowship under Grant No. DGE-2140743. Any opinion, findings, and conclusions or recommendations expressed in this material are
those of the author(s) and do not necessarily reflect the views of the National Science
Foundation.

AI Disclosure Statement: The authors implemented generative AI (\texttt{Claude} Opus 5.5) in the use of text refinement suggestions and optimization of code written by the authors.
\end{acknowledgments}
\software{
\texttt{matplotlib} \citep{Hunter2007},
\texttt{numpy} \citep{harris2020array},
\texttt{scipy} \citep{2020SciPy-NMeth},
\texttt{pandas} \citep{reback2020pandas}.
}
\appendix 
    \section{Scattering Geometry} \label{sec:scat_geo}
Consider the lab frame (unprimed) and the encounter frame (primed). The encounter frame defines the velocity changes relative to the initial relative velocity vector, and one must then project the velocity changes in the encounter frame to that of a chosen lab frame. In the lab frame, we choose to orient our unit vector in the x-direction ($\hat{e_x}$) along the direction of our test object's original velocity, so that $\Vec{v} = v \hat{e_x} + 0 \hat{e_y} + 0\hat{e_z}$.

In the encounter frame, a natural basis can be defined using the unit vector of the relative velocity $\hat{V}$ and impact parameter vector $\hat{b}$. Then, we can write $ \Vec{ \Delta v} = \Delta v_V \hat{V} + \Delta v_b \hat{b}$. To write the relative velocity unit vector in the lab frame
\begin{equation}
    \hat{V} = \frac{V_x}{V} \hat{e_x} - \frac{v_{fy}}{V} \hat{e_y} - \frac{v_{fz}}{V} \hat{e_{z}}\,.
\end{equation}
With $V_x = v-v_{fx}$. Defining $\alpha= V_x/V$ and $\beta = -\sqrt{1-\alpha^2}$ and introducing a variable $\phi$ as the azimuthal angle around $\hat{e_x}$ such that $\hat{e_\rho} = \cos(\phi) \hat{e_y} + \sin(\phi) \hat{e_z}$, we see
\begin{equation}
    \hat{V} = \alpha \hat{e_x} + \beta \hat{e_\rho}
\end{equation}

To construct the basis in the primed frame, first define $\hat{e_x}' = \hat{V}$. Then, define another basis vector in the plane spanned by $\hat{e_x}$ and $\hat{V}$. Guessing a form of $\hat{e_y}' = c \hat{e_x} + d \hat{e_\rho}$, one can determine $c=-\beta$ and $d=\alpha$ suffices as another basis vector. Thus, 
\begin{equation}
    \hat{e_y}' = -\beta \hat{e_x} + \alpha \cos(\phi) \hat{e_y} + \alpha \sin(\phi) \hat{e_z}
\end{equation}

\begin{figure*}
    \centering
    \hspace{0.07\linewidth}\includegraphics[width=0.9\linewidth]{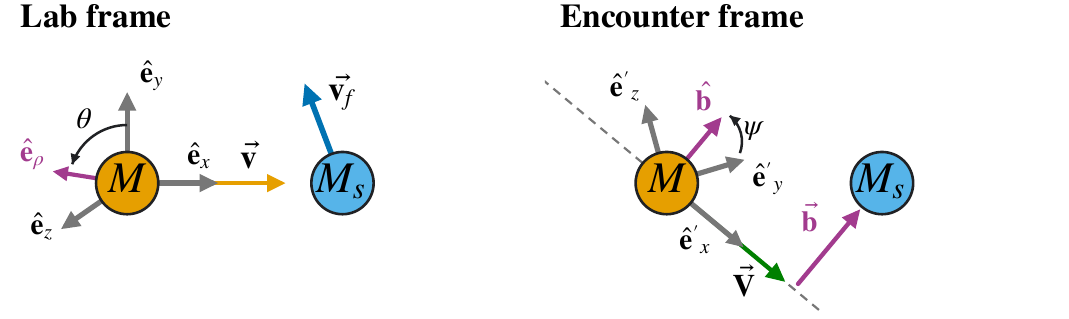}
    \caption{Scattering geometry schematic showing both the lab (unprimed) and encounter (primed) frames, along with the relevant coordinate and axis definitions.}
    \label{fig:frame_fig}
\end{figure*}

To construct $\hat{e_z}'$, take the cross product of $\hat{e_x}' \times \hat{e_y}'$ using the lab basis components, finding $\hat{e_z}' \equiv \hat{e_x}' \times \hat{e_y}' =  0 \hat{e_x} -\sin(\phi) \hat{e_y} + \cos(\phi) \hat{e_z} $. To then express Eq.~\ref{eq:dv_enc} in terms of the primed basis, introduce an azimuthal angle in the scattering frame $\psi$ such that $\hat{b} = \cos(\psi) \hat{e_y}' + \sin(\psi) \hat{e_z}'$. The geometry of both primed and unprimed frames is shown in Figure~\ref{fig:frame_fig}. We can then simplify to write
\begin{equation}
\begin{gathered}
     \Vec{\Delta v}= \bigg[\Delta v_V \alpha - \Delta v_b \beta \cos(\psi)\bigg] \hat{e_x} 
    + \bigg[[\Delta v_V \beta  + \Delta v_b \alpha \cos(\psi)] \cos(\phi) - \Delta v_b  \sin(\psi) \sin(\phi) \bigg] \hat{e_y} \\
    + \bigg[[\Delta v_V \beta  + \Delta v_b \alpha \cos(\psi)] \sin(\phi) + \Delta v_b  \sin(\psi) \cos(\phi) \bigg] \hat{e_z}
\end{gathered}
\end{equation}

One sees that
\begin{equation}
     \dvpar = \Delta v_x =  \Delta v_V \alpha - \Delta v_b \beta \cos(\psi) \label{eq:dvpar_b_V_psi}
\end{equation}
For the magnitude of the perpendicular component, we have
\begin{equation}
    \Delta v_\perp^2 \equiv  \Delta v_y^2 + \Delta v_z^2 = (\Delta v_V \beta + \Delta v_b  \alpha \cos(\psi))^2 + \Delta v_b^2 \sin^2(\psi) \label{eq:dvper_b_V_psi}
\end{equation}
Which, as one would expect, is independent of $\phi$, although the quantities $\dvpar$ and $\dvper$ depend on $\psi$. We have that the magnitude of the velocity kick yields
\begin{equation}
    \Delta v_{||}^2 + \Delta v_\perp^2 = \Delta v_V^2 + \Delta v_b^2\,.
\end{equation}
Thus, the magnitude of the kick is invariant under our coordinate change.
\section{Kick Distribution Derivation} \label{sec:drd_der}
From Eq.~\ref{eq:rate_measure}, first make the variable change from $\{v_{fx}, v_{fy}, v_{fz} \} \to \{ V_x, V, \theta \}$.  
Then, a Jacobian transformation results in ${\rm d}^3 v_f = V {\rm d}V \,{\rm d}V_x\,{\rm d} \theta$. Assuming $f$ is isotropic 
\begin{equation}
    {\rm d}\Gamma = f(V, V_x) V^2 b \,{\rm d}V\,{\rm d}V_x\,{\rm d}b\,{\rm d} \psi\,{\rm d}\theta\,.
\end{equation}
 We have then that
\begin{equation}
\begin{gathered}
        \frac{{\rm d}^2\Gamma}{{\rm d} \dvpar  {\rm d} \dvper} = 2\pi \int f(V, V_x) V^2 b   \delta(\dvpar - \dvv \alpha + \dvb \beta \cos(\psi)) \delta(\dvper - \sqrt{\dvv^2 + \dvb^2 - \dvpar^2} )
        \,{\rm d}V\,{\rm d} V_x\,{\rm d}b\,{\rm d}\psi
\end{gathered}
\end{equation}
Where we have already integrated over $\theta \in [0,2\pi)$. Then, let $\psi_*$ to be where the left $\delta$ function is zero. This requires
\begin{equation}
   (\dvb \beta )^2 - (\dvv \alpha - \dvpar)^2 \ge 0\,,
\end{equation}
and so defining $\Omega^{1/2}  =  \sqrt{\dvb^2 \beta^2 - (\dvv \alpha - \dvpar)^2} $ we have
\begin{equation}
    \int_{0}^{2\pi} \delta(\dvpar - \dvv \alpha + \dvb \beta \cos(\psi)) {\rm d}\psi = \frac{2 \Theta(\Omega)}{\sqrt{\Omega}}\,,
\end{equation}
where $\Theta$ is the Heaviside function. We adopt the convention with $\Theta(0)=0$. To evaluate the remaining delta function, it is convenient to change variables. Let $u = b/b_0$, with $b_0 = G(M+M_{\rm s})/V^2$. Then, performing this change of variables results in
\begin{equation}
    \frac{{\rm d}^2\Gamma}{{\rm d} \dvpar  {\rm d} \dvper} = 4 \pi G^2 (M+M_{\rm s})^2 \int \frac{u f(V, V_x)}{V^2} \frac{\Theta(\Omega)}{\sqrt{\Omega}}  \delta(\dvper - \sqrt{\dvv^2 + \dvb^2 - \dvpar^2})
         \,{\rm d}V\,{\rm d}V_x\,{\rm d}u\,.
\end{equation}
To perform the integral involving the second $\delta$-function, define $u_*$ to be the value of $u$ where the remaining $\delta$-function argument is zero. Then $u=u_*$ implies
\begin{equation}
    u_* = \sqrt{ \frac{4\mu^2 V^2}{\dvpar^2 + \dvper^2}-1}\,.
\end{equation}
The argument of the root requires $4\mu^2 V^2 \ge \dvpar^2 + \dvper^2$. For $u\in [u_\mathrm{min}, u_\mathrm{max}]$,
\begin{equation}
    \int_{u_\mathrm{min}}^{u_\mathrm{max}} u \delta(\dvper - \sqrt{\dvv^2 + \dvb^2 - \dvpar^2})\,{\rm d}u = \frac{4\mu^2 \dvper V^2}{(\dvpar^2 + \dvper^2)^2} \Theta(u_*-u_\mathrm{min}) \Theta(u_\mathrm{max}-u_*)\,.
\end{equation}

Enforcing these restrictions on $\Omega$ we have 
\begin{equation}
    \Omega_*= \Omega\big|_{u=u_*} = (  \dvpar^2 + \dvper^2 - \frac{(\dvpar^2 + \dvper^2)^2}{4 \mu^2 V^2}) (1-\frac{V_x^2}{V^2}) - ( -\frac{\dvpar^2 + \dvper^2}{2\mu V} \frac{V_x}{V} - \dvpar)^2\,.
\end{equation}
Then, we have

\begin{equation}
\begin{gathered}
        \frac{{\rm d}^2 \Gamma}{{\rm d}\dvpar {\rm d} \dvper} = 16\pi G^2 M_{\rm s}^2 \frac{\dvper}{(\dvpar^2 + \dvper^2)^2} \int {\rm d}V {\rm d}V_x \frac{f(V, V_x)}{\sqrt{\Omega_*}}  \Theta(\Omega_*) \Theta(u_*-u_\mathrm{min}) \Theta(u_\mathrm{max}-u_*) \Theta(4\mu^2 V^2 - \dv^2)
\end{gathered} \label{eq:kick_dist_wintegrals}\,.
\end{equation}

 Without loss of generality, take the perpendicular kick direction to be aligned with the y-axis, so that $\Delta \Vec{v} = \dvpar \hat{e_x}+ \dvper \hat{e_y} +0 \hat{e_z}$. Now, take a general relative velocity $\Vec{V} = V_x \hat{e_x}+ \sqrt{V^2-V_x^2} \cos(\theta) \hat{e_y} + \sqrt{V^2 - V_x^2} \sin(\theta) \hat{e_z}$. Then,
\begin{equation}
    \Delta \Vec{v} \cdot \Vec{V}  = -\frac{\dv^2}{2\mu} \to \dvpar V_x + \dvper \sqrt{V^2-V_x^2}\cos(\theta) + \frac{\dv^2}{2\mu}=0\,.
\end{equation}
Define a quantity $h=h(\theta)$ so that
\begin{equation}
    h(\theta) \equiv  \Delta \Vec{v} \cdot \Vec{V}  +\frac{\dv^2}{2\mu} = \dvpar V_x + \dvper \sqrt{V^2-V_x^2}\cos(\theta) + \frac{\dv^2}{2\mu}\,.
\end{equation}
Notice that the integration of $h(\theta)$ yields
\begin{equation}
    \int_{0}^{2\pi} \delta(h(\theta)){\rm d}\theta = \frac{2}{\sqrt{\dvper^2 (V^2-V_x^2) - (\frac{\dv^2}{2\mu}+ \dvpar V_x)^2}} = \frac{2}{V \sqrt{\Omega_*}}\,.
\end{equation}
But then, notice that
\begin{equation}
    \int f \delta(h) {\rm d}^3 v_f = \int f \delta(h(\theta)) V {\rm d}V {\rm d}V_x {\rm d}\theta = 2\int \frac{f}{\sqrt{\Omega_*}} \,{\rm d}V\, {\rm d}V_x
\end{equation}
Here, the Heaviside functions have been omitted, which is permitted in the case of large $u_\mathrm{max}$ and $u_\mathrm{min}=0$. Further, $2\mu V > \dv$ from Eq~\ref{eq:scat_iden1} and the Cauchy-Schwarz inequality. The definition of $\kappa$ allows us to rewrite the scattering condition as
\begin{equation}
    h = \dv (\kappa - \Delta \Vec{v} \cdot \Vec{v_f}/\dv)\,.
\end{equation}
Inserting this allows for the expression of the integral in Eq.~\ref{eq:kick_dist_wintegrals} as
\begin{equation}
   \int \frac{f(V, V_x)}{\sqrt{\Omega_*}}{\rm d}V {\rm d}V_x =  \frac{1}{2\dv} \int f(v_f) \delta(\kappa - \Delta \Vec{v}\cdot \Vec{v_f}/\dv) {\rm d}^3 v_f\,.
\end{equation}
This is a two-dimensional integral, and can be interpreted as intersecting the three-dimensional field star velocity space with the plane where $ \Delta \Vec{v}\cdot \Vec{v_f}/\dv = \kappa$ is enforced. This intersection then has minimum distance $|\kappa|$ from the origin, and conveniently traces out a disk to integrate over if the field star distribution is isotropic. A similar form and interpretation is derived in \cite{Goodman1983}. Generically, we can write Eq.~\ref{eq:kick_dist_wintegrals} instead as
\begin{equation}
    \frac{{\rm d}^2\Gamma}{{\rm d}\dvpar {\rm d}\dvper }= 8\pi G^2 M_{\rm s}^2 \frac{\dvper}{(\dvpar^2 + \dvper^2)^{5/2}} \int_{\Delta \hat{v}\cdot \Vec{v_f} = \kappa} f(v_f)\, {\rm d}A
\end{equation}
Where $dA$ is the area element on the plane in the allowed region where $|\kappa| \le v_\mathrm{esc}$. Then, one can introduce a radial coordinate in the disk to perform the area integral.

\section{Marginal Distributions}\label{sec:appendix_margs}
The fully explicit forms of the $\mathcal{F}$ and $\mathcal{G}$ functions are
\begin{equation} \label{eq:curly_F_G}
    \begin{gathered}
        \mathcal{F}(\dvper) = \frac{1}{8\mu^2 \sqrt{\dvpar^2 + \dvper^2}} + \frac{\dvpar v}{6 \mu  (\dvpar^2 + \dvper^2)^{3/2}} + \frac{\dvpar^2 v^2}{10(\dvpar^2 +\dvper^2)^{5/2}} - \frac{1}{3 (\dvpar^2 + \dvper^2)^{3/2}} \frac{G\Mbh}{r}\,,\\
        \mathcal{G}(\dvpar) = \frac{G \Mbh}{r} \frac{2 \dvpar^3 +3 \dvpar \dvper^2}{3 \dvper^3 (\dvpar^2 + \dvper^2)^{3/2}}- \frac{\dvpar}{8\mu^2 \dvper \sqrt{\dvpar^2 + \dvper^2}}+ \frac{\dvper v}{6\mu (\dvpar^2 + \dvper^2)^{3/2}}- \frac{(2\dvpar^5 + 5 \dvpar^3\dvper^2)v^2}{30\dvper^3 (\dvpar^2 + \dvper^2)^{5/2}}\,.
    \end{gathered}
\end{equation}
\section{Diffusion Coefficients} \label{sec:appendix_dc}
For the velocity-space DCs, we have \citep{Merritt2013}
\begin{equation}
\begin{gathered}
\langle \dvpar \rangle = -16 \pi^2 G^2 M_{\rm s} (M+M_{\rm s})\ln (\Lambda) F_2(v)\,, \\
    \langle \dvpar^2 \rangle = \frac{32\pi^2}{3} M_{\rm s}^2 G^2 \ln(\Lambda) v \big[ F_4(v) +E_1(v) \big]\,,\\
    \langle \dvper^2 \rangle = \frac{32\pi^2}{3} M_{\rm s}^2 G^2 \ln(\Lambda) v \big[ 3 F_2(v) - F_4(v) +2 E_1(v) \big] \,.\\
\end{gathered}\label{eq:Diff_cos}
\end{equation}
Here, $F_n(v)$ and $E_n(v)$ are defined as 
\begin{equation}
    F_n(v) = \int_{0}^{v} \bigg{(}\frac{v_f}{v}\bigg{)} ^n f(v_f) {\rm d} v_f\,\,,\,\,\, E_n(v) = \int_{v}^{\infty} \bigg{(}\frac{v_f}{v}\bigg{)} ^n f(v_f) {\rm d} v_f
\end{equation}
To calculate the local second-order DC in action space, one has
\begin{equation}
     \langle \dJ^2 \rangle = \frac{J^2}{v^2} \langle \dvpar^2 \rangle + \frac{1}{2} (r^2 - \frac{J^2}{v^2}) \langle \dvper^2 \rangle \,,\,\,\,\, \langle \dE^2 \rangle = v^2 \langle \dvpar^2 \rangle \,.
\end{equation} 

In the NSC discussed in this work, when computing $\langle \dE^2 \rangle$ for the object in the radial part of its orbit, $E_1 \sim 0$ as the bounds of integration are between $v$ and $\vesc$. Correspondingly, the bounds on $F_4$ are approximately between $0$ and $\vesc$. Then, we may write
\begin{equation}
\begin{gathered}
        F_4 \approx C_\gamma \left( \frac{G\Mbh}{r}\right)^{\gamma-3/2} \int_0^{\vesc} \left( \frac{v_f^4}{\vesc^4}\right)\left( 1-\frac{v_f^2}{\vesc^2}\right)^{\gamma-3/2}\,{\rm d} v_f = C_\gamma \left( \frac{G\Mbh}{r}\right)^{\gamma-3/2} \frac{\vesc}{2}  \int_0^{1} \nu^{3/2} \left( 1-\nu \right)^{\gamma-3/2}\,{\rm d} \nu  
\end{gathered}
\end{equation}
Where the last integral is the beta function $B(5/2, \gamma-1/2)\equiv B_\gamma$. Then, we can express $\langle \dE^2 \rangle$ as in Eq.~\ref{eq:dE2_loc_approx}.
\section{Orbit Averaging}
To average a general quantity $X$ over a Keplerian orbit, 
\begin{equation}
    X_o = \frac{2}{P} \int_{r_p}^{r_a} \frac{X}{v_r}\,{\rm d} r \label{eq:orb_avg}\,.
\end{equation}
For the case of the highly eccentric orbit, one can approximate that $v_r\approx \vesc$ on the radial portion of the orbit, and $v\sim0$ at apocenter.
\bibliographystyle{aasjournalv7}
\bibliography{biblio.bib}

\end{document}